\documentclass[11pt]{article}
\usepackage{fullpage}
\usepackage{xcolor}
\usepackage[colorlinks=true, citecolor=green!60!black,linkcolor=blue]{hyperref}
\usepackage[left=1in,right=1in,top=1in,bottom=1in]{geometry}
\usepackage{booktabs}
\usepackage{amssymb}
\usepackage{amsmath}
\usepackage{framed}
\usepackage[ruled]{algorithm2e}
\usepackage[export]{adjustbox}
\usepackage{mathtools}
\usepackage{cleveref}
\usepackage{amsmath, amsthm,csquotes}
\usepackage{amsthm}
\usepackage{fancyhdr}
\usepackage{subcaption}
\usepackage{makecell}
\usepackage{natbib}
\usepackage{csquotes}
\setcitestyle{authoryear}

\SetAlFnt{\small}
\SetAlCapFnt{\small}
\SetAlCapNameFnt{\small}
\SetAlCapHSkip{0pt}
\IncMargin{-\parindent}

\newcommand{\supp}{\operatorname{supp}}
\newcommand{\majority}{\text{maj}}
\newcommand{\minority}{\text{min}}

\newcommand{\R}{\mathbb{R}}
\newcommand{\U}{\operatorname{\emph{U}}}
\newcommand{\RS}{\operatorname{\emph{RS}}}
\newcommand{\Pois}{\operatorname{Pois}}
\newcommand{\CS}{\operatorname{CS}}
\newcommand{\Rec}{\operatorname{Rec}}
\newcommand{\argmax}{\operatorname{arg\,max}}

\newcommand{\E}{\mathbb{E}}

\newcommand{\ta}{\tilde{a}}
\newcommand{\tg}{\tilde{g}}
\newcommand{\coldstart}{\textsc{Cold Start}}
\newcommand{\recommendation}{\textsc{Recommendation}}
\newcommand{\MTurk}{\texttt{MTurk}}

\newtheorem{theorem}{Theorem}[section]

\newtheorem{remark}[theorem]{Remark}

\newtheorem{proposition}[theorem]{Proposition}

\newtheorem{example}[theorem]{Example}
\newtheorem{definition}[theorem]{Definition}

\graphicspath{img/}

\makeatletter
\newcommand{\subalign}[1]{%
  \vcenter{%
    \Let@ \restore@math@cr \default@tag
    \baselineskip\fontdimen10 \scriptfont\tw@
    \advance\baselineskip\fontdimen12 \scriptfont\tw@
    \lineskip\thr@@\fontdimen8 \scriptfont\thr@@
    \lineskiplimit\lineskip
    \ialign{\hfil$\m@th\scriptstyle##$&$\m@th\scriptstyle{}##$\hfil\crcr
      #1\crcr
    }%
  }%
}
\makeatother

\makeatletter
\renewenvironment*{displayquote}
  {\begingroup\setlength{\leftmargini}{0.2cm}\csq@getcargs{\csq@bdquote{}{}}}
  {\csq@edquote\endgroup}
\makeatother

\usepackage{parskip}
\title{Recommending to Strategic Users}
\author{Andreas Haupt\thanks{Massachusetts Institute of Technology, \url{haupt@mit.edu}} \and Dylan Hadfield-Menell\thanks{Massachusetts Institute of Technology, \url{dhm@csail.mit.edu}} \and Chara Podimata\thanks{UC Berkeley  \& Massachusetts Institute of Technology, \url{podimata@mit.edu}}}
\begin{document}
\maketitle
\begin{abstract}
Recommendation systems are pervasive in the digital economy. An important assumption in many deployed systems is that user consumption reflects user preferences in a static sense: users consume the content they like with no other considerations in mind. However, as we document in a large-scale online survey, users \emph{do} choose content strategically to influence the types of content they get recommended in the future. 

We model this user behavior as a two-stage noisy signalling game between the recommendation system and users: the recommendation system initially commits to a recommendation policy, presents content to the users during a cold start phase which the users choose to strategically consume in order to affect the types of content they will be recommended in a recommendation phase. We show that in equilibrium, users engage in behaviors that accentuate their differences to users of different preference profiles. In addition, (statistical) minorities out of fear of losing their minority content exposition may not consume content that is liked by mainstream users. We next propose three interventions that may improve recommendation quality (both on average and for minorities) when taking into account strategic consumption: (1) Adopting a recommendation system policy that uses preferences from a prior, (2) Communicating to users that universally liked (\enquote{mainstream}) content will not be used as basis of recommendation, and (3) Serving content that is personalized-enough yet expected to be liked in the beginning. Finally, we describe a methodology to inform applied theory modeling with survey results.
\end{abstract}
\section{Introduction}\label{sec:introduction}
Recommendation systems---the technology at the heart of online platforms---are inextricably entwined with our everyday lives. From movies (e.g., Netflix, Hulu) to short blogs (e.g., TikTok, Twitter, Mastodon) and e-commerce (e.g., Amazon), people turn to these recommendation systems to select entertainment, information, and products. For example, a recent study by \citet{gomez2015netflix} revealed that $80\%$ of the approximately $160$ million hours of video streamed on Netflix were \emph{recommended} by the service's recommendation system. In light of this, countless papers have been written exploring how recommendation algorithms learn about users' preferences, and how users interact with recommendations.

From what we understand, users and recommendation systems exist in a feedback loop:
recommendation systems supply a user with recommendations, and in response, the user decides whether and how to engage with this recommended content. From this engagement, the system receives information about the user's preferences in the forms of \emph{explicit} feedback---e.g., likes, comments, shares, and \emph{implicit} feedback---e.g., information about which content the user elected to engage with. The system, in turn, uses its inferences about the user's preferences to supply the user with future recommendations.  

Many models treat the user as a passive participant in this feedback loop, assuming that users simply consume what they like without regard for their ability to influence the content they will see in the future. However, others posit that users may actively try to affect what the platform learns about their preferences in order to \enquote{curate} their feeds based on the content they want to see more of \citep{simpson2022tame}. If users truly engage in curation, this has significant implications for how platforms learn users' preferences, because users' behavior in response to recommendations may be strategic, rather than simply maximizing enjoyment of content. And if platforms misrepresent users' preferences, there might be grave dangers of unfairness arising if minorities have differential incentives to engage in curation. This motivates the following two questions:
\begin{displayquote}
\emph{\textbf{Question 1}: Are everyday users of online platforms aware of this feedback loop when they are curating their feeds? And if yes, what types of actions (if any) do they take in response?}

\emph{\textbf{Question 2}: Can we provide theoretical insights on the harms that arise in recommendation to strategic users? And are there any interventions that can provably ameliorate these harms?}
\end{displayquote}

\subsection{Empirical Contributions}\label{sec:contributions}
We answer Question 1 affirmatively via a large-scale survey run on Amazon Mechanical Turk ($\MTurk$), in which we asked participants about their consumption behavior on TikTok, the popular short video hosting platform (see \Cref{sec:survey}). The majority of our respondents (nearly $60\%$) are not only aware of this feedback loop, but they explicitly and actively curate their feeds in order to influence the platform to serve them content of a particular type in the future. We find that users build \enquote{folk theories} about, for example, how the platform creates associations among different content and types of users. Some of the folk theories described by respondents attempt to explain why the platform puts a particular type of content into their feed, and include the following: 
\begin{displayquote}
\emph{\enquote{I feel that Tiktok continues to put these in my feed because I almost always get sucked into watching them. That tells the algorithm that I like them, even though I am mostly just using them for background noise and have seen most of them before.}}

\emph{\enquote{TikTok probably puts this into my feed because it is close to the eating challenges videos. I am
interested in that, so it probably ties the two together. Also I will watch those videos.}}

\emph{\enquote{I respond well and engage with funny videos. The AI learns from that.}}

\emph{\enquote{I think because I liked a video once of this type of content. I believe by me liking the video, the
algorithm thought I would like to see more videos like that one.}}
\end{displayquote}
Based on these folk theories, the participants in our survey reported taking specific actions in order to influence the content they want the recommendation system to put on their feed in the future. The actions they reported are sometimes efforts to reinforce the algorithm's perception that they belong in a particular user type. 
\begin{displayquote}
\emph{\enquote{I also like stuff just to see more
of that type stuff evn though I don’t like it. LIke soemtimes if my content gets to dark I try to like animal videos and comedy more to get off the darker content for a bit.}} [sic]

\emph{\enquote{Sounds odd but sometimes I will click on something once, get out and click it again then search for it if it’s a topic I like that I realize I haven’t been seeing. I don’t know if that actually works but it seems to.}}

\emph{\enquote{Currently, I am cognizant of what category of video I think material falls under. I am careful to watch completely videos that fall under the correct category (even if I am not interested in that particular video). I am careful to skip over videos from the "wrong" categories.}}
\end{displayquote}
Other times, they are actions that hide specific parts of their identity from the algorithm.
\begin{displayquote}
    \emph{\enquote{I actually have an acute interest in religion but I prefer my religious
engagement to be somewhere less frivolous than tik tok.}}

\emph{\enquote{i don’t follow any tiktok hikers or anything like that. tiktok doesn’t know i have an interest in
it.}} [sic]

\emph{\enquote{Again I don't look up vidoes about retail because I'm at work most of the time and don't want to see work related stuff on my free time.}}
\end{displayquote}
There are many quotes similar to the ones listed above. We list them in \Cref{sec:quotes}.

We use this affirmative answer to question 1 a launching pad for the theoretical part of our work. Our goal is to present a (game-theoretic) model that accounts for user behavior and to study the societal effects of user strategizing.

\subsection{Theoretical Contributions}\label{sec:theory-contributions}

Our first contribution is to present a formal framework (\Cref{sec:model}) for the interaction between the recommendation systems and users who strategically consume content to affect the recommendations they get at a later stage. Henceforth, we refer to this behavior as \enquote{users strategizing}.

The interaction between the principal (aka recommendation system, $\RS$) and the agents (aka users, $\U$) proceeds in $3$ phases. In the first phase, $\RS$ \emph{commits} to \emph{recommendation policy} $g$ mapping \emph{consumption frequencies} $q \in Q:= \mathbb{N}^d$ to \emph{content} that is to be served in the future, $x \in X:= [d]$, where $d$ denotes the number of different content types. $\U$ gets to observe the recommendation policy $g$. Due to the principal's commitment power, our setting resembles a \emph{Stackelberg game}. 

The second phase is known as the $\coldstart$. The $\coldstart$ refers to the phase during which the system is presenting a lot of (exploration) content to users with limited consumption history in order to learn their preferences. In other words, it is an inference step where the system suggests a variety of different content, collects the user interaction, and ultimately recommends new content based on the users' observed consumption patterns. We model the $\coldstart$ phase to do exactly this: a user of \emph{type} $\theta \in \Theta$ arrives, they make a \emph{consumption plan} $a_\theta \in A:= [0,1]^d$ about the type of content they want to consume, and ultimately consume said content with frequencies $q \sim \pi(a_\theta)$, where $\pi(\cdot)$ is the \emph{consumption realization function}. We assume that the lever that the users can pull in order to strategize is to alter their consumption plan $a_\theta$. We study two equilibrium notions for $\RS$: the first one is the \emph{Stackelberg Equilibrium (SE)}, where $\RS$ optimizes policy $g$ accounting for $\U$ strategizing in the $\coldstart$. In the second notion, \emph{Naive Platform Equilibrium (NPE)}, $\RS$ optimizes $g$ thinking that $\U$ is acting \emph{myopically} in the $\coldstart$.

We remark that the $3$-phase interaction is a simplified view of how recommendation systems work in practice. In reality, $\RS$ keeps learning the users' preferences in the third phase too, so the learning and recommending phases are more intricately intertwined. Our separation in a cold start and a recommendation phase is with the goal to highlight forces that likely exist also in highly dynamic interactions on a recommendation system.

Our second contribution (\Cref{sec:challenges}) is to identify the effects of $\U$ strategizing. We show that in our model, when $\RS$ is playing in the NPE, $\U$'s best response is to strategize. We show that when the user population is split into \emph{minority} and \emph{majority} users in the NPE the minority users will overconsume intra-group preferred content type, \emph{even if} they do not actually like the particular piece of content that is currently served to them. This is in line with what our participants reported:
\begin{displayquote}
    \emph{\enquote{I make sure to interact things that are specific to content types I want to see, even if I don’t really love the content of that specific video.}}
\end{displayquote}
\begin{displayquote}
    \emph{\enquote{If there was a private mode, I would use that to search things that I wouldn't want recommended to me. Stuff that I like, but stuff that I wouldn't want to clog my feed.}}
\end{displayquote}
In an NPE, minority may also not consume mainstream content at all, as a way to better signal their minority group identity. This, of course, is rather unfair for them, since they do not consume content they actually like. We next show that these behaviors may reduce welfare, and that strategizing has a negative externality on users within a user group. 

Given these undesirable effects, in \Cref{sec:interventions} we present three interventions for $\RS$ to ameliorate them. The first intervention, \emph{over-representing minorities}, focuses on the recommendation policy and the properties it satisfies. This can be implemented by increasing the probability that the content suggested to a user comes from the minority-preferred content types. The second intervention, \emph{automatic incognito mode}, focuses on the information that $\RS$ collects about $\U$. Such a mode would allow $\U$ to consume content but simultaneously signal to $\RS$ that they do not want this content consumption to change their recommendation. The final intervention that we consider, \emph{changing the $\coldstart$}, focuses on carefully engineering the content that is served in the $\coldstart$ so that it is more informative and it improves welfare.

Coming full circle, the last contribution of this paper is putting forth a new paradigm for combining applied modeling and theoretical research. Since the Machine Learning and the Computer Science communities are increasingly being interested in questions that pertain to human behavior in the interaction with modern technological systems, we believe that instead of treating theoretical models separately from user surveys we should instead use user surveys to inform our modeling choices. We outlined our process for doing this, along with our most important qualitative findings, in \Cref{sec:survey}.

\subsection{Related Work}

\paragraph{Recommendation Systems Design.} There has been a vast literature on recommendation system design, mostly from a practical standpoint. For the theoretical framework that we put forth, we took into account considerations about recommendation systems that arise in practice (e.g., the user $\coldstart$ problem of recommending to users about whom very limited consumption history has been observed \cite{safoury2013exploiting, zheng2017identification}). We study the harms that can arise from user strategizing---both to minorities and to all users---and propose interventions that platforms can enact to mitigate them. In this sense, our work fits in the literature on aligning recommendation systems with user preferences~\cite{stray2021you,kleinberg2022challenge}. Note here that our work highlights the negative effects of users strategizing \emph{even if} the recommendation system is benevolent and perfectly aligned with the user's interests.  

\paragraph{User perception of algorithms.} Our work (and especially our qualitative survey study) is closely related to a literature in the field of Human Computer Interaction regarding how users believe online platforms work, the folk theories they build about their algorithms, and how this shapes their content consumption behavior on said platforms \cite{eslami2016first, lee2022algorithmic, klug2021trick}. Close to our work, \citet{simpson2022tame} study whether users can actively influence the algorithm to represent particular parts of their identity. Contrary to these studies, the purpose of our survey was to inform our modeling assumptions.

\paragraph{Signalling.} From the economic theory literature, our work is related to \emph{signalling} and, more broadly, strategic communication. In a signalling game (see e.g., \cite{spence1978job,crawford1982strategic}), an informed sender needs to send information to an uninformed receiver. In particular, close to our work are papers on the role of noise in signalling \cite{blume2007noisy,landeras2005noisy,de2011noisy}, as the consumption realization function $\pi(\cdot)$ may be seen as a noisy channel via which the sender communicates with the receiver regarding their preferences. Furthermore, the Poisson consumption model, which is the main instantiation of our general model of recommendation to strategic users, bears similarities to models of strategic communication with lying costs \cite{kartik2009strategic,deneckere2007optimal}. In all these works, conflicts of interest between the sender and receiver are a focus of stufy. However, in this paper we consider settings where the sender (i.e., the user) and receiver (i.e., the recommendation systemplatform) preferences are aligned, and we still identify negative effects of strategic behavior. As another difference, we also study receivers with commitment power, similar Stackelberg games, which have been used as the main modeling tool for problems in the next research area we highlight.

\paragraph{Pedagogy and Cooperative Inverse Reinforcement Learning} The signalling problem we consider also relates to the literature on cooperative inverse reinforcement learning (CIRL)\citep{hadfield2016cooperative,malik2018efficient}. In these, a human sender tries to signal their preferences to an autonomous agent through actions (also called \enquote{demonstrations}). In contrast to classical inverse reinforcement learning \citep{ng2000algorithms}, which assumes that these actions are taken according to the human's preferences, CIRL considers a model in which the human \enquote{teaches} the autonomous agent to achieve a long-term goal. Our model may be framed as a CIRL problem, where the autonomous agent is a recommendation system. In contrast to the classical focus of CIRL, however, we focus on environments with a large number of humans, and expose biases arising in such preference elicitation.

\paragraph{Learning with strategic data sources.}Finally, our work is related to the literature on \emph{learning with strategic data sources}, where a principal interacts with agents who strategize with the data they feed into the principal's algorithms in an effort to obtain better outcomes from the principal's algorithm. We can divide this literature into two broad threads. The first one adopts the \emph{principal}'s perspective and wishes to design learning algorithms who satisfy different forms of robustness to the strategizing of the agents. Some examples of the forms of robustness that have been considered are incentive-awareness (e.g., \cite{hardt2016strategic,dong2018strategic,chen2020learning,ahmadi2021strategic,ghalme2021strategic,levanon2021strategic}), truthfulness/strategyproofness (e.g., \cite{meir2012algorithms, cummings2015truthful, chen2018strategyproof, ball2019scoring, eliaz2019model}), performativity \cite{perdomo2020performative}, and causal inference (e.g., \cite{shavit2020causal,bechavod2021gaming}). The second thread adopts \emph{society's} perspective and studies the fairness implications of algorithms that are designed with the robustness to strategizing goal in mind (see e.g., \cite{hu2019disparate, milli2019social, braverman2020role, bechavod2022information}). Our study follows the second strand of research, points out implications of algorithms for users, and proposes interventions that may improve the outcomes for users. On a technical level, our model of strategic recommendation differs from most previous models in that it most closely resembles an \emph{unsupervised} learning problem, as opposed to a supervised one.
\section{Model \& Preliminaries}\label{sec:model}
\begin{figure}[htbp]
\includegraphics[width=\textwidth]{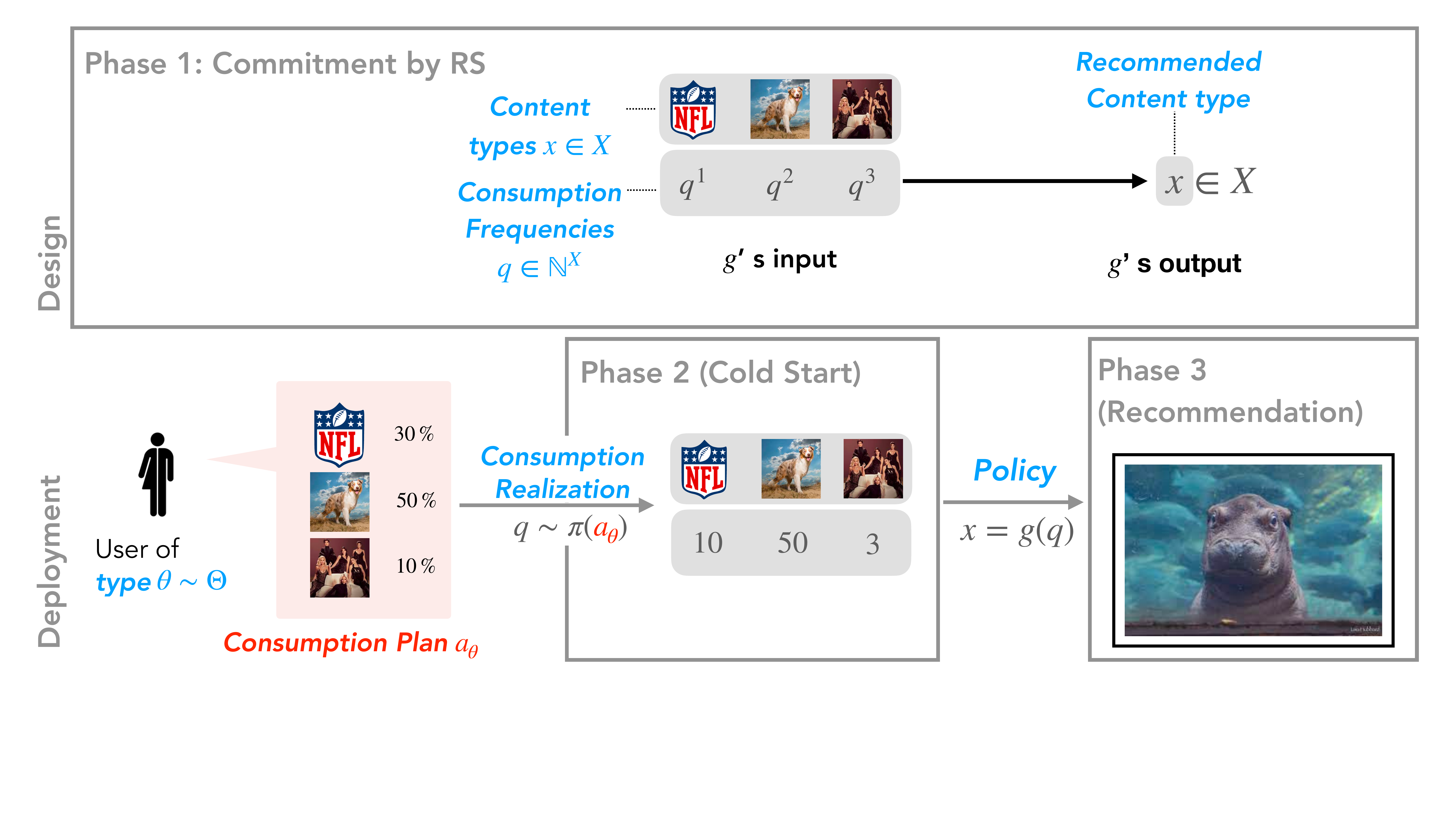}
\caption{Pictorial Representation of our Model: In Phase 1, $\RS$ commits to policy $g$ mapping realized consumption frequencies to content. During Phase 2, the user of type $\theta$ is exposed to $3$ types of content: $x^1 = \texttt{NFL}$, $x^2 = \texttt{dog}$, and $x^3 = \texttt{Kardashians}$. The user then makes a plan to consume $a_\theta = (30\%, 50\%, 10\%)$ of the $x^1, x^2, x^3$ respectively. Ultimately, the user ends up consuming contents $(x_1, x_2, x_3)$ for a total of $q = (10, 50, 3)$ times. The consumption profile $q = (10, 50, 3)$ is then observed by $\RS$ who uses it as input on their committed recommendation policy $g$ to decide which content to serve to the user in Phase 3, the $\recommendation$. In this example, and because the user consumed mostly the animal-related content in the $\coldstart$ phase, the content served in $\recommendation$ is animal-related. To be precise, the picture in the $\recommendation$ phase of the figure is of Fritz, the baby hippo from the Cincinnati Zoo.}
\label{fig:model-sketch}
\end{figure}

In this section, we first introduce our general framework, called \emph{strategic recommendation}, which we later instantiate for the \emph{Poisson consumption} model. A pictorial and simplified version of our model can be found in \Cref{fig:model-sketch}.

\paragraph{Conventions} Uppercase letters refer to sets (e.g., $Q$) and matching lowercase letters, e.g., $q, q'$, to representative elements. Letters $\mu, \nu$ refer to probability distributions, and $\varepsilon, \delta >0$ to real numbers. Subscripts (e.g., $a_\theta$) refer to user types and superscripts (e.g., $a^j$) refer to content types.

\subsection{The Strategic Recommendation Problem}
We consider an interaction between a recommendation system $\RS$ and users $\U$, where users choose content to consume from a set $X = \{1,2, \dots, d\}$ recommended to them by $RS$. Users are of stochastic, heterogeneous types $\theta \in \Theta \coloneqq [0,1]^d$ (with $\theta^j$ corresponding to the probability that a user enjoys content of type $j \in [d]$), distributed according to $\mu \in \Delta(\Theta)$.\footnote{A model with a finite number of preference profiles and content types resembles low-rank models used in recommendation systems \cite{recht2011simpler}.} 

The interaction between $RS$ and $U$ --- which we term the \emph{Strategic Recommendation} problem --- occurs in three phases and is specified in Protocol~\ref{protocol:strategic-recommendation}. In the protocol, we denote by $a_\theta \in A := [0,1]^d$ the \textit{consumption plan} of all users with type $\theta$, which specifies for each content type $j \in [d]$ the probability $a_\theta^j$ that a user of type $\theta$ engages with content type $j$ when exposed to it. For the game theorist, this means that we assume symmetric strategies, which we assume for simplicity throughout the article. Correspondingly, we define $q \in Q:= \mathbb{N}^d$ to be the users' \textit{observed} consumption frequencies, i.e., $q^j$ specifies the number of times a user consumed content $j$ when exposed to it.
\begin{algorithm}[htbp]
        \SetAlgoNoLine
        \LinesNumbered
        \SetAlgorithmName{Protocol}{}{}
        \begin{enumerate}
        \item $\RS$ commits to a \emph{recommendation policy} $g \colon Q \to X$ mapping \emph{realized consumption} $q \in Q$ to content served in the future, $x \in X$. 
        \item {\bf [\coldstart]} $\U$ of type $\theta$ makes a \emph{consumption plan} $a_\theta \in A$ for the content they want to consume. $\RS$ observes the \emph{realized consumption}  $q \sim \pi(a_\theta)$. We call $\pi \colon A \to \Delta(Q)$ the \emph{consumption realization function}. 
        \item {\bf [\recommendation]} $\U$ is served content $g(q)$ which they choose to consume based on their utility.
        \end{enumerate}
        \caption{Interaction Protocol in Strategic Recommendation}
        \label{protocol:strategic-recommendation}
\end{algorithm}
In words, the three-phase interaction starts with $\RS$ committing on a recommendation policy. It is followed by a $\coldstart$ phase where $\U$ consumes content and $\RS$ is learning about $\U$'s preferences. It concludes with the $\recommendation$ phase, in which $\RS$ suggests new content to $\U$ based on the preferences $U$ has declared and $\RS$'s recommendation policy. $\U$ can strategize only during the $\coldstart$ phase in order to affect the content that they see in the $\recommendation$ phase.

To capture users' differing goals in the $\coldstart$ versus the $\recommendation$, $\U$ (and thus $\RS$) have different utility functions in the two phases. We assume the utility functions across the phases are additively separable, so $\U$ and $\RS$'s overall utility functions, $u_{\U}, u_{\RS} \colon Q \times X \times A \times \Theta \to \R_+$ are:
\begin{equation}
\begin{split}
u_{\U}(q, x, a; \theta ) &= u_{\U}^\text{CS} (q, a ; \theta) + u_{\U}^\text{Rec} (x ; \theta) \\
u_{\RS}(q, x, a ; \theta) &= u_{\RS}^\text{CS} (q , a; \theta) + u_{\RS}^\text{Rec} (x ; \theta)
\end{split}\label{eq:U-utility}
\end{equation}
In words, $u_{\U}(q,x, a ; \theta)$ corresponds to the utility that an agent of type $\theta$ obtains when they consume content at the frequencies specified by $q \in Q$ in the $\coldstart$ phase after having made a consumption plan $a$, and then they get served content of type $x \in X$ in $\recommendation$ phase. Many of the challenges in strategic recommendation arise even absent incentive conflicts, so we will assume that the preferences of $\U$ and $\RS$ are aligned, i.e., $u_{\U}^{\CS}=u_{\RS}^{\CS}$ and $u_{\U}^{\Rec}=u_{\RS}^{\Rec}$. 

We consider two behavioral models for $\RS$. The first one is the Stackelberg Equilibrium (SE) and it assumes that $\RS$ adapts to the strategizing of $\U$ and computes the optimal recommendation policy \emph{anticipating} that $\U$ will best respond to it. Note that the computation of an SE also assumes that the platform has a correct estimate (or full knowledge) of the prior distribution of user types $\mu \in \Delta(\Theta)$.

\begin{definition}\label{def:BNS-equilibrium}
A pair $\left((a^*_\theta)_{\theta \in \Theta}, g^*\right)$ is a Stackelberg Equilibrium (SE) for $\U$ of type $\theta$, if $\RS$ maximizes
\[
g^* \in \arg \max_{g\colon Q \to \Delta(X)} \,  \E_{\subalign{q &\sim \pi\left(a^*_\theta\right)\\\theta &\sim \mu}} \left[ u_{\RS}(q, g(q), a^*_\theta; \theta)\right] 
\]
where $a^*_\theta$ is the best-response of $\U$ given recommendation policy $g^*$, i.e.,
\[
a^*_\theta \in \arg \max_{a_\theta \in A} \E_{q \sim \pi(a_\theta)} \left[ u_{\U} \left(q, g^*(q), a_\theta ; \theta \right)\right]
\]
We will also use $a = (a_\theta)_{\theta \in \Theta}$ as a shorthand notation for a profile of consumption plans. 
\end{definition}
We contrast an $\RS$ that plays the SE to one that does not account for the strategizing of the users when they choose their consumption plans and takes the realized consumptions in the $\coldstart$ phase at face value. We refer to the latter as one that acts in a Naïve Platform Equilibrium (NPE). To distinguish between SE and NPE strategies in our notation, $\bullet^*$ signifies the SE strategy and $\tilde{\bullet}$ signifies the NPE strategy.

\begin{definition}[Naïve Platform Equilibrium (NPE)]\label{def:BNE}
A triple $\left((\tilde a_\theta)_{\theta \in \Theta}, \tilde g, \tilde \mu\right)$ is a Naïve Platform Equilibrium (NPE) if the $\RS$ maximizes
\begin{equation}
\tg \in \arg \max_{g:\; Q \to \Delta(X)} \E_{\subalign{q &\sim \pi(a'_\theta)\\\theta &\sim \tilde\mu}} \left[ u_{\RS} \left( q, g(q), a_\theta'; \theta\right)\right]\label{eq:npebr}
\end{equation}
where $a'_\theta$ is a \emph{myopic} (i.e., only in the cold start) best-response of the user of type $\theta$ to $\tg$, i.e.,
\[
a'_\theta \in \arg \max_{a_\theta \in A} \E_{q \sim \pi(a_\theta)} \left[ u_{\U}^{\CS}\left(q, a_\theta ; \theta \right) \right].
\]
The agent in reality reacts optimally to $\tg$:
\[
\ta_\theta \in \arg \max_{a_\theta \in A} \E_{q \sim \pi(a_\theta)} \left[ u_{\U}\left(q, \tg(q), a_\theta ; \theta \right) \right]
\]
We assume that the distributions $\mu$ and $\tilde \mu$ are compatible, i.e., the consumption distribution from users using $\tilde a_\theta$ under $\mu$ is the same as the one under $a'_\theta$ under $\tilde\mu$. Technically, the random variables $q \sim \pi(\tilde a_\theta)$ where $\theta \sim \mu$ and $q' \sim \pi(a'_\theta)$, where $\theta \sim \tilde\mu$ have the same distribution.
\end{definition}
The definition of an NPE relies on a misspecification of the recommendation system; the $\RS$ assumes that agents are myopic, while in reality, they are not. While, in principle, many distributions of user types $\tilde \mu \in \Delta(\Theta)$ are possible in an NPE, our main instantiation of Strategic Recommendation (the Poisson consumption model we present below) implies a unique choice of $\tilde \mu$.

\begin{remark}[User vs Platform Knowledge]
The SE and NPE differ in their assumption on the knowledge that the $\RS$ possesses regarding the user behavior. In both equilibrium notions, users know their own type $\theta$, the $\RS$'s recommendation policy $g^*$ (resp. $\tg$), the consumption realization function $\pi$, their utility functions $u_{\U}^{\CS}$ and $u_{\U}^{\Rec}$ and the spaces $Q$, $A$, and $X$. In an SE, the $\RS$ knows in addition the type distribution $\mu \in \Delta(\Theta)$.
\end{remark}

\subsection{The Poisson Consumption Model}
\begin{figure}[htbp]
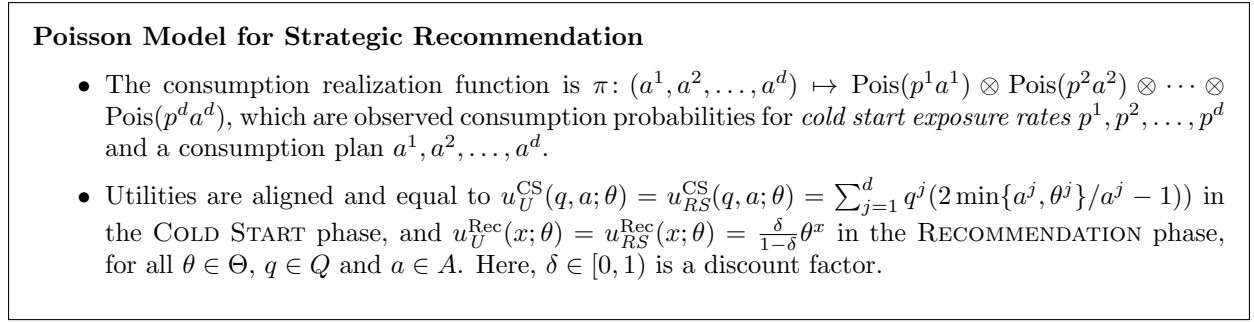

\begin{framed}
\small\textbf{Poisson Model for Strategic Recommendation}
\begin{itemize}
    \item The consumption realization function is $\pi \colon (a^1, a^2, \dots, a^d) \mapsto \Pois (p^1 a^1) \otimes  \Pois (p^2 a^2) \otimes \cdots \otimes  \Pois (p^d a^d)$, which are observed consumption probabilities for \emph{cold start exposure rates} $p^1, p^2, \dots, p^d$ and a consumption plan $a^1, a^2, \dots, a^d$.
    \item Utilities are aligned and equal to
    $u_{\U}^{\CS} (q,a;\theta) = u_{\RS}^{\CS} (q,a;\theta) = \sum_{j=1}^d 
    q^j (2\min\{a^j, \theta^j\}/a^j - 1))$ in the $\coldstart$ phase, and
    $u_{\U}^{\Rec} (x;\theta) = u_{\RS}^{\Rec} (x;\theta)  = \frac{\delta}{1-\delta}\theta^x$ in the $\recommendation$ phase, for all $\theta \in \Theta$, $q \in Q$ and $a \in A$. Here, $\delta \in [0,1)$ is a discount factor.
\end{itemize}
\end{framed}
\caption{Instantiation of our Strategic Recommendation Model in the Poisson Setting}\label{fig:poisson}. 
\end{figure}

We instantiate the Strategic Recommendation setting for the \emph{Poisson consumption model} for $\U$. To micro-found this modeling  choice, we start by assuming that users do not recall their previous interactions with the $\RS$. Games with no recall have been discussed in game theory, compare \cite[Section 11.1.3]{osborne1994course}. For each of the \emph{content types} $j \in [d]$, there is a probability $\theta^j$ that the users will like the content they are served. The $\RS$ repeatedly, in the $n \gg 0$ rounds of the $\coldstart$ phase, shows the user random content of type $j$ with probability $p^j$. A user repeatedly chooses whether or not to consume the served content with probability $1-1/n$. As the user does not recall any previous interaction with the content, they can only form a (potentially mixed) action distribution mapping the type of the content and the agent's preference to consumption or not. Together with the distributional assumption, this yields a reduced-form model outlined in \Cref{fig:poisson}. Consumption of \emph{liked} content yields a utility of $1$, consumption of \emph{non-liked} content a utility of $-1$, and \emph{non-consumption} a utility of $0$. For a large number $n \gg 0$ of rounds, Poisson approximation \cite[p.302]{billingsley2008probability} leads to consumption that is Poisson-distributed. 

We now explain the closed form of the $\coldstart$ utility for $\U$ of type $\theta$ stated in Figure~\ref{fig:poisson}. Fix a content type $j \in [d]$. We distinguish two cases. If $\U$'s consumption plan is such that $a^j \leq \theta^j$, $\U$ derives the highest utility (while generating the same consumption counts observable by the platform) by consuming all the content they like and no other content. This yields a utility of $q^j$. Otherwise, the optimal consumption $\theta^j < a^j$ any piece of content that is consumed liked with probability $\theta^j/a^j$ and not liked with probability $(1-\theta^j /a^j)$.\footnote{This is a consequence of Bayes' rule. $\Pr[\text{like} | \text{consume}] = \Pr [ \text{consume} | \text{like}] \cdot \Pr [ \text{like}] /\Pr[ \text{consume}]$ and $\Pr[\text{like}]  = \theta^j$, $\Pr[\text{consume}] = a^j$. Also, $\Pr[\text{consume}|\text{like}] = 1$, which is optimal for the user.} This means that the utility in this case is: $q^j (1 \cdot \frac{\theta^j}{a^j} + (-1) \cdot ( 1 - \frac{\theta^j}{a^j})) = q^j \left( 2\frac{\theta^j}{a^j} - 1 \right)$. Putting everything together, the \emph{expected} utility in the $\coldstart$ phase for the Poisson model is:
\begin{equation}
\E_{q \sim \pi(a)}[u_{\U}^{\CS} (q, a; \theta)] = \sum_{j=1}^d p^ja^j(2 \min\{\theta^j, a^j\}/a_j - 1)= \sum_{j=1}^d p^j(2 \min\{\theta^j, a^j\} - a^j).\label{eq:uniquemax}
\end{equation}
Note that this function has a unique maximizer at $a^j = \theta^j, j =1 , 2, \dots, d$. In particular, a naïve platform can infer from this a distribution of types justifying myopic optimality.

The closed form for the recommendation phase utility is the utility of the user repeatedly consuming content that they like of the type of content that is recommended, with a discount factor of $\delta$, assuming that there is no discounting within the initial round. 

\begin{proposition}[NPE for the Poisson Model]\label{prop:single-peaked-poisson}
In the Poisson model, $\argmax_{a} \E_{q \sim \pi(a)}[u_{\U}^{\CS} (q, a; \theta)]$ has a unique maximizer. In addition, an NPE can be characterized by a \emph{tuple} $((\tilde a_\theta)_{\theta \in \Theta}, \tg)$ such that 
\[
\tg \in \arg \max_{g:\; Q \to \Delta(X)} \E_{\subalign{q &\sim \pi(a'_\theta)\\\theta &\sim \tilde \mu}} \left[ u_{\U} \left( q, g(q), \tilde a_\theta; \theta\right)\right] \quad \text{and} \quad 
\ta_\theta \in \arg \max_{a_\theta \in A} \E_{q \sim \pi(a_\theta)} \left[ u_{\U}\left(q, \tg(q), a_\theta ; \theta \right) \right],
\]
where $\tilde \mu$ is the distribution of $\ta_\theta$. 
\end{proposition}
Note that this definition involves, as in SE, only two components: A choice of recommendation policy $\tilde g$, and consumption plan choices $(\ta_\theta)_{\theta \in \Theta}$. The inferred distribution of user types $\tilde \mu$ in an NPE is forced to be the distribution of the consumption plan $\ta_\theta$. This can be interpreted as the action choices in the Poisson model being \enquote{truthful} for myopic agents, i.e., $\ta_\theta  = \theta$ for all $\theta \in \Theta$. This simplifies a NPE compared to the general case since the distribution $\tilde \mu$ is implied by other equilibrium variables.

\begin{proof}
The unique maximization is direct from \Cref{eq:uniquemax}. Aligned incentives allow us to replace $\RS$ utilities by $\U$ utilities. The choice of $\tilde \mu$ is compatible with $\mu$ as, because of \Cref{eq:uniquemax}, $\tilde{a}_\theta = \theta$ holds for all $\theta \in \Theta$. Because of this equality constraint and the fact that $\pi \colon [0,1]^d \to \Delta(\mathbb N^d), a \mapsto \Pois(p^1a^1) \otimes \Pois(p^2a^2) \otimes \cdots \otimes \Pois(p^da^d)$ is a bijective function, there cannot be any other compatible distributions $\tilde \mu$.
\end{proof}
\section{The Challenges of Strategic Users in Recommendation Systems}\label{sec:challenges}
In this section, we discuss how $\U$ in the Poisson consumption model \emph{strategically distort} their myopically optimal consumption plans in efforts to change the types of content suggested to them by $\RS$. Subsequently, we discuss the negative effects of this behavior. We start by defining \emph{myopic optimality}.

\begin{definition}
    A user type $\theta \in \Theta$ \emph{acts myopically optimally} if $a_\theta^j = \theta^j$ for all content types $j = 1, 2, \dots, n$. If a user type does not act mypically optimal, we say they \emph{distort their consumption}.
\end{definition}
We first observe that the Poisson consumption model is able to produce consumption distortions: Myopic optimality, or the assumption that users reveal their preferences, need not hold.
\begin{theorem}[Existence of Consumption Distortions]
In the Poisson consumption model, if the best response $g$ of the platform \eqref{eq:npebr} serves at least two types of content, no agent is exactly indifferent between the served types of content, and all types are interior, $\theta^j \in (0,1)$ for all $j = 1, 2, \dots, d$, then there is a scalar $\overline{\delta} \in (0,1)$ such that for any $\delta \ge \overline{\delta}$, myopic optimality is not an NPE.
\end{theorem}
Note that this result does not imply that all types of users need to distort their consumption. In fact, in the main running example we introduce below, we will see that a group of users (the \enquote{majority}) does not distort their consumption, as they foresee that their distortion costs them too much in the cold start phase to be justified by the gains in being recommended their favorite content.

\begin{proof}
    Assume for contradiction that $a_\theta = \theta$ was part of an NPE. As the discount factor $\delta$ is uniform across agent types, the optimal function $g$ in an SE and NPE for myopically optimal user consumption will be independent of $\delta$. Observe that, in the Poisson consumption model, the function $[0,1]^d \to \Delta(X), a \mapsto g(\pi(a))$ as Poisson tails are continuous in the rate parameter. As there are two different types of content that are served, there is a type of content $x \in X$ such that the set of realized consumption profiles $g^{-1}(\{x\})$ is neither empty nor the full set of realized consumption, $\emptyset \subset g^{-1}(\{x\})\subset Q$. By the structure of the Poisson consumption model---in particular that the Poisson probability mass function is single-peaked---it holds that $g^{-1}(x)$ is closed: If $q, q' \in g^{-1}(x)$ and $q \le q'' \le  q'$ in the component-wise partial order, then $q'' \in g^{-1}(x)$. Because of this, we can choose $x$ such that there is $j$ such that $0 \in g^{-1}(\{x\})$. Let $\theta \in \Theta$ be a type such that $\argmax_{j=1, 2, \dots, d} \theta^j = x$. As $\theta \in (0,1)^d$, each individual user gains a constant probability of being recommended their favorite content type if they do not consume content $j$ at all, $a_\theta^j =0$. As no agent type is indifferent between the two served content types, this leads to a constant gain of utility, which we call $C \delta/(1-\delta)$. The loss of utility from not consuming content $j$ is, in expectation, at most $\theta^j$, as types are upper-bounded by $1$. For $\overline{\delta} > {\theta^j}/(C + \theta^j)$, this deviation is profitable.
    \end{proof} 

Next, we establish a particular NPE in an example; it explains how to reason about NPEs, but also highlights problematic user incentives that recommendation to strategic users might lead to.

\begin{example}[Majority and Minority]\label{ex:main}
Consider a population of majority and minority users, $\Theta \coloneqq \{\majority, \minority\}$. The minority, as the name suggests, is smaller than the majority. We will use a particular size of a minority, which we will motivate later, $\mu(\minority) = 1/(1+e^{-(1-\varepsilon)/3})$. 
There are $d=3$ content types, which we call $\text{majority-preferred}$, $\text{mainstream}$ and $\text{minority-preferred}$. We will index them in this order, i.e., use $j=1$ for $\text{majority-preferred}$, $j=2$ for $\text{mainstream}$, and $j=3$ for $\text{minority-preferred}$. We assume that the types are $\theta_{\majority} = (1, 1-\varepsilon, 0)$, $\theta_{\minority} = (0, 1-\varepsilon, 1-\varepsilon)$. Hence, the majority is more likely to derive utility from majority-preferred content than mainstream content, but does not like any minority content. The minority is as likely to like mainstream content as minority-preferred content, but does not like majority-preferred content. 
There is an equal distribution of the three types of content in the $\coldstart$ exposure distribution, $p^1 = p^2 = p^3 = 1/3$. We will assume that $\delta$ is such that $\delta/(1-\delta) = 2$ makes the future twice as important as the present. We will also assume that $\varepsilon \in (0,\frac14)$.

Hence, there are two groups of users of different sizes, that have overlapping consumption profiles. We will show that in equilibrium, this overlap will disappear, and that the minority will distort its consumption. In particular, the rest of this example establishes that the following is an NPE:
\begin{align*}
g(x, y, z) &=\begin{cases}
\text{minority-preferred} & (x = 0 \text{ and } y = 0) \text{ or } z > 0\\
\text{majority-preferred} & \text{else.}
\end{cases}\\
a_{\majority} &= (1, 1-\varepsilon, 0)\\
a_{\minority} &= (0,0,1)
\end{align*}

First, observe that given the agent consumption choices, the platform recommendation policy $g$ is optimal in the sense of \eqref{eq:npebr}: For realized consumption profile $(x,y,z)$ such that $z>0$, it is certain that the user is a minority user, and hence should be recommended the minority-optimal content, which is $\text{minority-preferred}$. It is also certain that a realized consumption profile with $(x,y,z)$, $x > 0$ or $y > 0$ comes from a user that is a majority user. Note that under the consumption profile that the users chose, they appear symmetric in that the utility loss from mis-classification is the same for both groups. Hence, the recommendation system would like to maximize posterior probability of a user being either a minority or majority user. To show that the recommendation system's policy is optimal, we need to show that $\Pr [\theta = \minority | q = (0,0,0)] \ge 1/2$. By Bayes' law, this probability is equal to
\begin{multline*}
\frac{ \Pr[\theta = \minority] \Pr[ q_\minority = (0,0,0)]}{ \Pr[\theta = \minority] \Pr[ q_\minority = (0,0,0)] + \Pr[\theta = \majority] \Pr[ q_\majority = (0,0,0)]} 
\\
= \left(1 + \frac{\mu(\minority)}{1-\mu(\minority)} \frac{1 \cdot 1 \cdot e^{-1/3}}{e^{-1/3} e^{-(1-\varepsilon)/3}}\right)^{-1} = \left(1 + \frac{\mu(\minority)}{1-\mu(\minority)} e^{-(1- \varepsilon)/3}\right)^{-1} = \frac12
\end{multline*}
where the last line uses the choice of $\mu (\minority)$ from above.

Next, consider why the actions are optimal for the users. Majority users consume myopically optimally in $\coldstart$. Algebra shows that no deviation to $(1,y,0)$, where $y \in (1-\varepsilon, 1)$ benefits the majority. Consider now minority users. First, consuming any of the majority-preferred content can only increase the probability to be served majority content, yet reduces utility from the cold start phase. We show that for mainstream content, the minority user utility is decreasing in their consumption of mainstream content, and for minority-preferred content it is increasing. Note that in an NPE, by definition, the belief that the platform has about the user distribution is \emph{passive} (in the language of e.g., \cite{ichihashi2020online}), i.e., deviating users are infinitesimal in that it does not change the distribution of user preferences that the platform bases its recommendation on. We may hence evaluate the  value of a deviation by $g$ assuming that the recommendation policy is unchanged. We can evaluate the minority user probability from a consumption profile $(0, a^2, a^3)$ in closed form using \eqref{eq:uniquemax} and the form of the distribution above: $\frac13 (2 \min\{1-\varepsilon, a^2\} - a^2)) +  \frac13 (2 \min\{1, a^3\} - a^3)) + 2e^{-a^2}(1-e^{-a^3})$. Observe that $a^2 > 1-\varepsilon$ does not optimize the minority user's utility. In the other cases, we can rewrite it to
\[
\max_{a^2, a^3 \in [0,1]} \frac{a^2}{3}  +  \frac{a^3}{3} + 2(1-\varepsilon)e^{-a^2}(1-e^{-a^3}).
\]
Note that this is maximized at $(a^2, a^3) = (0,1)$. This completes the equilibrium characterization.
\end{example}
This NPE possesses three features. First, majority users do not distort their consumption; they consume exactly the content that they like, and get recommended their favorite content unless they do not end up engaging with the platform at all (i.e., $q = (0,0,0)$). Second, minority users over-consume content that is associated with their group, even if they do not like the particular piece of content that is served to them. They, hence, behave more \emph{stereotypically} than their actual identity. Respondents in our survey reported this type of distortion in our survey (see relevant quote at \Cref{sec:contributions}). Finally, minority users do not consume content of the mainstream at all. The main reason for this is that this allows them to be classified as minority in the future even if they end up not being served any of their favorite content in the $\coldstart$.

Both the inequities, but also the type of consumption distortion outlined here are undesirable. Users of the (statistical) minority are incentivized to act more stereotypically than they are. We see that these behaviors are bad for each member of the group, through the following statement. 
\begin{theorem}[Consumption Distortions have a Negative Intra-Group Externality]
   Assume that $\lvert \supp \mu \rvert =2$. Consider any NPE $(a,g)$ where some agent type $\theta \in \Theta$ does not play myopically optimally.  Consider the action profile for which $a_{\theta}' = \theta$, $a_{\theta'}' = a_{\theta'}$, $\theta' \in \Theta \setminus \{\theta\}$. Let $g'$ be a platform best response as in \eqref{eq:npebr}. Then, a user would be (weakly) better off, would all players of the same type not distort their consumption.
   \begin{equation}
   \E_{q \sim \pi(a_\theta)} [u_{\U} (q, g'(q), a_\theta; \theta)] \ge \E_{q \sim \pi(a_\theta)} [u_{\U} (q, g(q), a_\theta; \theta)].\label{eq:externality}
   \end{equation}
\end{theorem}
The intuition behind this result is that consumption distortions are making the platform estimate the preferences of the group as more pronounced than they actually are. In the example we saw above, the platform thinks that the minority likes all content of the minority-preferred content type, which is not the case. It is merely a result of the consumption distortion of the minority.
\begin{proof}
Note that a group will choose $a_\theta^j \neq \theta^j$ for some $j=1, 2, \dots, d$ only if this increases the probability of being served the favorite content of their group in comparison to playing truthfully, as the distribution of the consumption of the other group members does not change from $(a_\theta)_{\theta \in \Theta}$ to $(a_\theta')_{\theta \in \Theta}$. Note that, as we are comparing a user that consumes according to an equilibrium in both cases, their cold start phase utility does not differ in both cases. Only the probability of being recommended their content differs. We show that the probability of the recommendation system classifying the user to their group is higher if the other users are not distorting their consumption. (Note that this is enough, as, if the consumption profile $a_\theta$ has a different maximum than $\theta$, this only can increase the utility of the user in case the other agents of this type are truthful.) As the type space is binary, it is the case that the misreporting happens in a direction to increase the difference from other groups' consumption plans. Consider, for example, that $a_\theta^j < \theta^j < a_{\theta'}^j$ for $\supp \mu = \{ \theta, \theta'\}$ and some content type $j = 1, 2, \dots, d$. As Poisson distributions are ordered in first-order stochastic dominance, for any $(q, q^{-j}) \in Q$, if the recommendation system assigns a user the preferred content of $\theta'$ if everyone plays $a_\theta$, it does so if the user plays $\theta$. Conversely, the set of realized consumption profiles that are assigned $\theta$'s favorite content is weakly larger if $\theta$ plays $\theta$ as opposed $a_\theta$. In particular, under the action $a_\theta$, the user's probability of being assigned their preferred content must be weakly larger. The case of $a_{\theta'}^j < \theta^j  <  a_\theta^j $ is proved in a similar way.
\end{proof}
Lastly, we return to the running example, and show that without strategizing, the welfare would be higher, while differently distributed.
\begin{example}[Strategizing in an NPE Reduces Welfare]\label{ex:welfare}
   We return to the binary example we analyzed earlier. We show that a solution without algorithmic recommendation leads to close to optimal welfare: The recommendation policy that just recommends mainstream content, and does not personalize, leads to no consumption distortions by any of the users, and leads to welfare of $1-\varepsilon$ for each user, as opposed to $1$ for all the users that are correctly classified in the NPE that we considered. It is also truthtelling. For $\varepsilon$ small (and an accordingly sized minority), a non-personalized recommendation yields arbitrarily close to optimal welfare, and does not lead to wasteful consumption distortions.

   Not recommending based on any of the data of the $\coldstart$, however, requires knowledge of the distribution of types among the users, a point that we will further investigate in \Cref{sec:posthoc}.
\end{example}
\section{Interventions}\label{sec:interventions}
The last section showed various negative outcomes that arise from the recommendation process with strategic users. We now turn to interventions that $\RS$ can take to ameliorate these problems. Note that $\RS$ can improve on almost all the objectives it desires if it had an accurate prior on the user types and committed to a recommendation policy based on that.
\begin{theorem}
The SE solution gives weakly higher welfare than the NPE solution.
\end{theorem}
\begin{proof}
Let $(a,g)$ be an NPE. Observe that the policy $g$ is feasible, and that the SE maximizes true user welfare. Hence, the welfare from an SE $(a', g')$ must be at least that of $(a,g)$. 
\end{proof}
The difference between an SE and an NPE lies in what the recommendation system bases its recommendation on. In an NPE, it takes the data as given at face value, and might base recommendations on, for example, realized consumption from stereotypical consumption plans. This might, in particular, lead to an NPE with less exposure of minority users with content they like.

Moving from an NPE to an SE in a realistic recommendation scenario is challenging, as in many recommendation systems, there is little data from users that is not affected by strategizing. This is in particular the case in social media applications that do not regularly query users for feedback independent of content consumption decisions. Still, as we have seen in our survey, many users claim to give such explicit signals of preference for content, which is not guided by short-term consumption decisions: Liking pages, following creators, and tapping explicit buttons to say that one would like to see less of something. The crucial difference between explicit feedback on content and data that is observed in consumption is that users do not need to make trade-offs between consuming content they like and their future recommendations. Data for recommendation based on explicit feedback may be used without reasoning about user strategizing in its interpretation. 

We will consider three types of interventions that improve the outcomes of NPEs and can be applied with only limited knowledge on the distribution of preferences in the user population: over-representing minorities in type distributions (\Cref{sec:posthoc}), making commitments about ignoring of data for recommendations (\Cref{sec:infodesign}) and designing the cold start distribution (\Cref{sec:coldstart}). 

\subsection{A Recommendation Choice Intervention: Over-Representing Minorities}\label{sec:posthoc}

We consider first an intervention that over-represents users from a particular user type in the recommendations. We start by defining the rescaling of a distribution.
\begin{definition}[Distribution Rescaling]\label{def:rescaling}
    Let $\nu \in \Delta(\Theta)$ be a probability distribution. For $\theta \in \Theta$ and $\alpha \in [1, 1/\nu(\theta)]$, call $\nu_{\theta, \alpha}$ the probability distribution with masses $\nu_{\theta, \alpha} (\theta) = \alpha \nu (\theta)$, and $\nu_{\theta, \alpha} (\theta') = \frac{1-\alpha \nu(\theta)}{1-\nu(\theta)}\nu (\theta')$.
\end{definition}
We will now consider an intervention into the algorithm that scales the probability in observed consumption frequencies $a_\theta$ up for some group. Hence, the algorithm assumes that minorities are a bit larger than they are in the data.

To this end, we call a recommendation policy $g$ an $(\alpha, a_\theta)$-rescaled best response if it best responds \eqref{eq:npebr} to $\tilde \mu_{\theta, \alpha}$, where $\mu_{\theta, \alpha}$ is the distribution rescaling from Definition~\ref{def:rescaling}. We consider the consequences of rescaling for a particular user group. We say that $(a,g)$ constitute an $(\alpha, a_\theta)$-rescaled NPE if $a$ is a best response to $g$ and $g$ is an $(\alpha, a_\theta)$-rescaled best response to $a$.

\begin{proposition}
Consider a binary type distribution $ \supp \mu  = \{\theta, \theta'\}$, and let $(a,g)$ be an NPE in which user type $\theta$ distorts consumption. Let $\alpha \in (1, \mu(\theta))$ be a scale-up parameter.  Then, there are actions $a_\theta'$ and $a_\theta''$ that, respectively, yield at least as high $\coldstart$ and $\recommendation$ phase utility than under $(a,g)$ if the platform $(\alpha, a_\theta')$- (resp. $(\alpha, a_\theta'')$)-best responds.
\end{proposition}
This proposition shows, that a minority may improve in their utility in both phases from over-representation in a model.
\begin{proof}
Observe that a higher probability weight in the observation increases, for each realized consumption profile $q\in Q$, the likelihood of being served the content of the over-represented group if type $\theta$ chooses a consumption plan $a_\theta'' = a_\theta$. This leads to the same cold start utility, and to at least as high recommendation phase utility.

Let $j \in \{1, 2, \dots, d\}$ be a content type that user type $\theta$ distorts consumption on. Assume that $(a_\theta')^j$ is infinitesimally closer to $\theta^j$ than $a_\theta$. By the definition of the Poisson consumption model, \eqref{eq:uniquemax}, this leads to a strictly higher cold start phase utility. As the probability masses of Poisson distributions vary continuously in the Poisson rate parameter, there is a change in the consumption that leads to at least as high posterior probability for agent type $\theta$ for all realized consumption profiles $q\in Q$. This means that the probability of being recommended content associated with group $\theta$ increases locally at $a_\theta$, which means that there is a consumption plan $a_\theta'$ that cold start utility.
\end{proof}
Note that while this intervention favors the group that is over-represented, it may decrease welfare, as other groups may be more likely misclassified and served content that they actually do not like.

In a deployed recommendation system, a rescaling would be most akin to a dataset rebalancing, an approach well known from supervised learning \cite{sklearn}. A recommendation system model would be trained with additional datapoints that are copies of a group (e.g., a minority) that should be over-represented. Our next intervention relates to disregarding some content consumption, and communicating this fact to users.
\subsection{An Information Design Intervention: Automatic Incognito Mode}\label{sec:infodesign}
In \Cref{ex:welfare}, we already saw an example of a recommendation policy that ignores all content consumption. This subsection considers interventions that commit to not using some of the consumption information.

In recommendation system models, one way to communicate this to users would be to automatically trigger an incognito mode when a user watches some type of content, meaning that the consumption (or not) of this content won't lead to a change in the recommendation. We say that a recommendation system makes a \emph{credible commitment to ignore} consumption of a content type if it restricts its maximization problem in \Cref{eq:npebr} to only those $g$ such that 
\begin{equation}
g(q^j , q^{-j}) = g((q^j)', q^{-j})\label{eq:ignoring}
\end{equation}
$\forall q^j, (q')^j \in [0,1]$ and $Q^{-j} \in [0,1]^{d-1}$. The \emph{ability} to make credible commitments improves welfare. 
\begin{proposition}[Commitments Around Non-Usage May Improve Welfare]
The ability to make credible commitments on ignoring some consumption weakly improves welfare, i.e., restricting the maximization problem in \eqref{eq:npebr} to \eqref{eq:ignoring} will not lead to any new equilibria that are welfare-dominated.
\end{proposition}
\begin{proof}
A restriction on the action set of a player, in this case the recommendation system, can never introduce new equilibria, in particular no new ones that are low in welfare.
\end{proof}
One consequence here is that ignoring content reduces wasteful user strategizing on this content.
\begin{proposition}[Effects of Automatic Incognito Mode]\label{prop:incognito}
Consider a Poisson consumption model with uninformative content type $j \in \{0,1, \dots, d\}$ in which a user $\theta$ distorts consumption. Then, there is an action profile $a'$ and a recommendation policy $g'$ that improves cold start phase utility.
\end{proposition} 
\begin{proof}
Define  $a'$ such that $(a_\theta')^{j} = \theta^j$ for all $\theta \in \Theta$. Clearly, this consumption plan weakly cold start utility. As there is a user that distorts consumption, there is at least one agent for which it strictly increases cold start utility.
\end{proof}
\begin{example}[Ignoring uninformative content]
We return to \Cref{ex:main}. Ignoring mainstream content in this example does not incentivize consumption distortions by $\theta_{\minority}$ and leads to significantly higher cold start phase utility for this group. This alternative leads to a consequence of less evidence. In fact, recommending minority-preferred content to $(0,0,0)$ may not happen in an NPE in which mainstream content is ignored, as can be seen as the recommendation system was indifferent in the case with consumption differences in mainstream content. Hence, the improvement in the cold start phase utility comes at a loss in the recommendation phase utility.
\end{example}
Uninformative content could take several shapes: e.g., popular content that is uniformly liked, or average content that is liked to a mediocre extent by all the user types.

\subsection{Information Gathering Interventions: Cold Start}\label{sec:coldstart}
A last intervention is to change the distribution of content that is served in the cold start phase of strategic recommendation, $\kappa$. This affects the recommendation policy through the consumption realization function $\pi_{\kappa}$. As the earlier two interventions, the possibility to change the cold start distribution only can improve welfare.
\begin{proposition}
    The ability to change the cold start weakly increases welfare.
\end{proposition}
Two features make a cold start induce low welfare: If Cold Start content is universally not liked by users---we will say \emph{has low qualtiy}---, i.e. $p$ puts a lot of mass on items for which $u_{\U}(\pi_p(a), x; \theta)$ is small for all $\theta \in \Theta$ and $a \in A$, then cold start phase utility will be low. If the content is uninformative, it may not be helpful in distinguishing users from it. Quality and informativeness are typically at a tradeoff. We merely show that if both quality and informativeness may be improved, then, in an NPE, welfare may be improved.

\begin{definition}[Quality Dominance]
We say that content $j = 1, 2, \dots, d$ \emph{quality-dominates} $j' = 1, 2, \dots, d$ if $\theta^j \ge \theta^{j'}$ for all $\theta \in \Theta$.
\end{definition}
\begin{definition}[Informativity Dominance]
We say that content $j \in \{1, 2, \dots, d\}$ informativity-dominates $j' \in \{1, 2, \dots, d\}$ if for any type $\theta, \theta' \in \Theta$, such that $\theta^j \le (\theta')^j$: $\theta^{j'} < \theta^{j} \le (\theta')^j < (\theta')^{j'}.$
\end{definition}
Informativity-dominant content accentuates differences in preferences among users. Replacing content by content that is of higher quality or more informative may improve utility in cold start respectively recommendation phase.
\begin{proposition}[Dominance in Cold Start Design]
Let $(a,g)$ be an NPE for cold start $\kappa$. If users consume content type $j'$ like content type $j$ under $(a,g)$, $(a,g)$ yields higher cold start utility under the cold start $\kappa$ that replaces $j$ by $j'$ if $j$ quality-dominates content $j'$ type $j'$. If $j'$ informativity-dominates content type $j'$, $a'$ such that $a'_\theta = \theta$ for all $\theta \in \Theta$, then the recommendation phase utility is increased.
\end{proposition}
\begin{proof}
The first statement follows from the fact that $j$ quality-dominates $j'$ that $(a,g)$ gives higher $\coldstart$ utility under $\kappa$ with $j$ replaced by $j'$ as a $\coldstart$ distribution compared to $\kappa$.

For the second statement, observe that the Poisson distributions under truthtelling have ordered rates. This means that the probability for each of the types to be classified correctly increases.
\end{proof}
In our example of a cold start distribution, content may not be ranked by quality and informativity.
\begin{example}[Cold Start in the Binary Model]
In the model introduced in \Cref{ex:main}, the three types of content are not ordered in any of the three dimensions: All three types of content are incomparable in quality, and mainstream is less informative than the other two types of content. In offering more mainstream content, the recommendation system trades off the ability to provide users with good recommendations in the future with cold start phase utility.
\end{example}
This concludes the three interventions we propose for strategizing. We next present our qualitative study testing some of the modelling assumptions we made in our modelling, and user behavior and assumptions in their interaction with TikTok's recommendation system.
\section{Survey}\label{sec:survey}
In this section we outline the most important findings from a large-scale survey of TikTok users. Our survey was run on Amazon Mechanical Turk (short \texttt{MTurk}) during December 2022 and January 2023. We solicited responses from $100$ participants from the USA regarding their interactions with content on TikTok. 

The basic statistics (demographics and usage) of our survey participants are reported in \Cref{appendix:survey-participants-stats}. The participant pool is \emph{gender balanced}, mostly \emph{white}, mostly \emph{between the ages $30-59$ years old} and \emph{diverse} in terms of content that they primarily consume on TikTok. The majority of the participants reported using TikTok daily, for more than a year, and for less than $2$ hours daily. 
The age distribution of respondents matches the demographics of $\MTurk$ workers, and skews \emph{older} than the average TikTok users (see e.g., \cite{TikTokdemographics}). 

Before the main survey whose results are reported here, we ran several exploratory pilot surveys of 10 respondents that allowed to extract areas of questions for our main survey. (We report our final survey questions in~\Cref{appendix:survey}). We then deployed the survey on $\MTurk$ with 118 responses, of which 100 were complete and considered in our analysis. Lastly, we qualitatively analyzed the resulting data. We describe our methodology in \Cref{appendix:methodology}. For most of our survey questions, we solicited free-form text responses from the participants. We report some of the major findings below. Our qualitative data analysis resulted in codes, which are given in \Cref{appendix:codes}

\subsection{Incognito Mode}

\begin{figure}[htbp]
\includegraphics[width=.43\linewidth,valign=t]{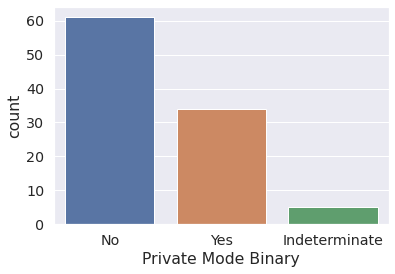}
\hspace{1cm}
\includegraphics[width=.46\linewidth,valign=t]{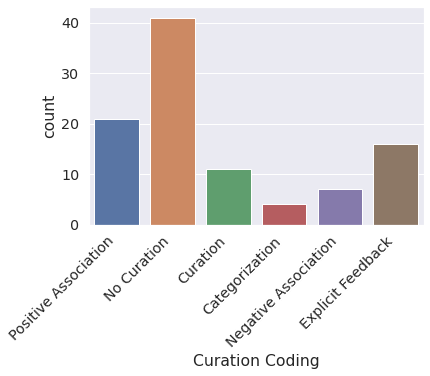}
\caption{\textbf{Left:} Responses on whether they would change their behavior on the platform, if TikTok were to offer an incognito mode. \textbf{Right:} Responses on whether they do anything to curate their feed.}
\label{fig:curation-privacy}
\end{figure}
Our first research question concerns whether, and why, users would change their behavior would TikTok use a private mode that does not record answers (see \Cref{fig:curation-privacy}, left). Such a private mode would have consequences for privacy, but would also mean that the recommendation system cannot adapt based on content consumed in the past.

Around $30\%$ of the participants responded that not only would their use of the platform change, but they would significantly expand on the content types that they consume. Respondents mention three categories of content they would seek out more under a private mode: (i) they would engage more with content they currently do not want to be associated with (e.g., \enquote{embarassing} or \enquote{risky} content), (ii) they would engage more with content that they do not want to \enquote{clog} their feed, and (iii) they would try to explore new content types that they have not been exposed to. Reasons (ii) and (iii) can be related to strategic content communication as defined in this article, as users express concerns over the impact of their consumption on the composition of their future feed. 

Several respondents stated that they would not change anything in their consumption patterns, one of the most common reason they listed was that any change would result in them seeing less personalized (and hence, worse) content. These responses illustrate our point that users of online platforms \emph{actively} consider the ways in which they consume content so as to change the way their feed will be in the future. The exact breakdown based on our encoding is reported in \Cref{table:private-mode}. 

Several participants highlighted in their responses the privacy implications of a private mode. This category of participants responded that they were not concerned with how their data was to be used and that they would not change anything in their consumption patterns, since they \enquote{have nothing to hide}. This is in line with other studies on how concerned people are about their data from social media being used~\cite{madden2013teens}.

{\footnotesize
\begin{table}[htbp]
\centering
\begin{tabular}{lr}
\toprule
{\bf Incognito Mode Coding} &  {\bf Participant Count} \\
\midrule
no change: no reason                            & 45 \\
no change: less personalization                 & 14 \\
\midrule
change: engage with \enquote{avoided} content          & 10 \\
change: engage with \enquote{feed-clogging} content    & 9  \\
change: exploration increase                    & 9  \\
\midrule
other                                           & 8  \\
\bottomrule
\end{tabular}
\caption{Encoding and statistics of the Incognito Mode question.}
\label{table:private-mode}
\end{table}}

\subsection{Curation}

Our second question (reported in \Cref{fig:curation-privacy}, right) concerns whether users engage in behaviors to curate their TikTok feed. The majority of our participants (around $60\%$) responded that they \emph{do} take actions to influence their future feed. Among the actions that they take, $21\%$ make sure to consume content they like to see more of, and $17\%$ make sure to give explicit feedback to the algorithm through likes, follows, etc. Interestingly, there are also some participants who specifically give evidence that they consume content that they do \emph{not} like so as to \emph{not be categorized} as not liking this type of content. These responses provide more evidence to the hypothesis that users of online platforms do actively and consciously consume content in efforts of affecting the way that the platform chooses what content to serve to them in the future. 

\subsection{Association Scenarios}

Our final question concerns how participants choose content given a particular goal of changing a future recommendation. We presented the participants with $4$ different scenarios regarding the content they would consume, if they wanted the algorithm to show them more/less from a related content type. 

\begin{figure}[htbp]
\centering
\begin{subfigure}[c]{0.30\textwidth}
\includegraphics[width=\linewidth]{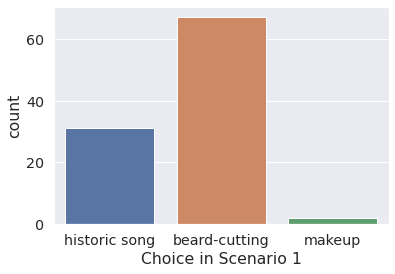}
\end{subfigure}
\hspace{.1\textwidth}
\begin{subfigure}[c]{0.30\textwidth}
\includegraphics[width=\linewidth]{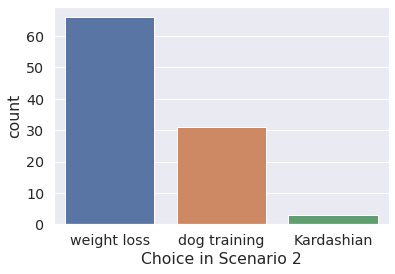}
\end{subfigure} \\\noindent
\centering
\begin{subfigure}[c]{0.30\textwidth}
\includegraphics[width=\linewidth]{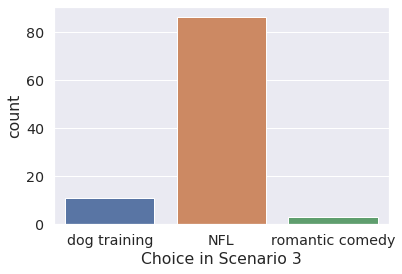}
\end{subfigure} 
\hspace{.1\textwidth}
\begin{subfigure}[c]{0.30\textwidth}
\includegraphics[width=\linewidth]{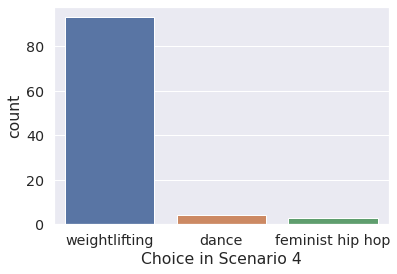}
\end{subfigure} 
\caption{Participants' aggregate responses to the $4$ different scenarios we presented them with. The detailed scenarios can be found in \Cref{appendix:survey}.}
\label{fig:scenarios}
\end{figure}
In the first and the second scenario (left and middle left of \Cref{fig:scenarios}), we asked the participants to choose the content they would consume in order to increase the sports content that the platform was recommending to them. In both cases, the majority of the participants associated \texttt{beard-cutting} and \texttt{weight loss} with sports. Even participants who chose the \texttt{song in a historic venue} and the \texttt{dog training} did report that they both have connections with sports (e.g., through the Super Bowl halftime performance, and the stereotypical image of a dog jogging with their guardian).  

In the third and fourth scenario (middle right and right in \Cref{fig:scenarios}), we asked the participants to choose the content they would consume if they wanted to see less make-up content. In both cases, the majority of the participants chose \texttt{NFL plays} and \texttt{weightlifting} as different from \texttt{make-up}. The agreement among participants was even stronger in these scenarios than in the first two scenarios.

\begin{figure}[htbp]
\centering
\includegraphics[width=.4\linewidth]{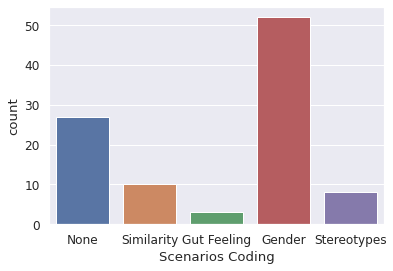}
\caption{Aggregate responses to how the participants made the associations in the \enquote{Scenarios} questions.}
\label{fig:scenarios-stats}
\end{figure}
Finally, in \Cref{fig:scenarios-stats} we report the aggregate statistics for the coding of the free-form answers of our participants regarding their decision-making in the scenario questions. We observe that more than $50\%$ of responders made the associations based on traditionally cis-gendered content, around $20\%$ based on stereotypes, and around $10\%$ based on content similarity.

This finding combined with the previous one regarding their \emph{intent} to curate their feed suggests that users do infer associations between different content types and use these correlations to affect the type of content the platform will serve to them in the future.
\section{Conclusions and Avenues for Future Work}\label{sec:conclusion}
There are several avenues for future work. First, several of our modeling assumptions could be relaxed, in efforts to better capture the intricasies of modern recommendation systems.

\paragraph{Conflicts of Interest.} Our model assumed that the platform is fully aligned with user interests and that it has the same utility function as the users. In reality, however, platforms have objectives that often diverge from user interests \cite{doctorow23}. An analysis of our model with a conflict of interest between the platform and users would allow for such a study. 

\paragraph{Estimation.} Our model assumes a perfectly Bayesian platform making inferences about the users' preferences based on their consumed content. However, platforms in reality are far from Bayesian. For example, popularity bias~\citep{abdollahpouri2019popularity} is a well-known failure of recommendation systems to account for exposure in their estimation of preferences. A model capturing features of real-world algorithms is an intriguing direction for future work. 

\paragraph{User Model.} Our model assumes that the users have perfect understanding of the recommendation system's policy. In reality,---and this is documented both in our survey and prior research \cite{eslami2016first}---users form \enquote{folk theories}; thus, when they best-respond, they do so with respect to their folk theory of the recommendation system, rather than the actual one. An interesting open question here concerns the effects of strategizing under different assumptions on rationality of users.

\paragraph{Supply Side Questions.} Several respondents in our qualitative study mentioned supply-side concerns in their responses when asked about why some part of their identities are not represented in their recommendations. The following two are examples of responses pointing to a folk theory about absence of content.

\begin{displayquote}
\emph{\enquote{I imagine there just aren't very many people in this field who are also inclined to make tiktok videos, or at least videos about their profession.}} [sic]

\emph{\enquote{I'm not sure if TikTok is not showing me this type of content. I'm starting to believe there just aren't TikTok creators who fall into this category? I see representations of all kinds of people who identify as LGBT, I just very, very rarely see anyone my age who is super feminine like me. There's tons of feminine gays on TikTok in their mid 20s and below, just not in my age range.}}
\end{displayquote}

These responses motivate the study of content creation incentives in recommendation systems. Intuitively, content creators (even if they share niche interests) may try to suppress them in order to strategize against the recommendation system that promotes more mainstream content, compare \cite{jagadeesan2022supply}.

Finally, the methodological paradigm adopted in this article can inform future applied theoretical research at the interface of Computer Science and Economics: Large-scale surveys of how users interact with algorithms can inform, and motivate, modeling choices.

\bibliographystyle{ACM-Reference-Format}
\bibliography{references}

\appendix
\newpage
\section{Methodology}\label{appendix:methodology}

In this section, we outline the methodology that we used for designing our large-scale survey and the analysis of its results. Our process for running the survey consisted of three phases.
\begin{enumerate}
\item In the first phase, we brainstormed different categories of questions (concept-driven approach). Our goal was to find the \enquote{correct} set of questions that would simultaneously achieve two goals. The first goal was to not be too leading (e.g., we did not want to ask explicitly whether they strategize with their content consumption to lead the algorithm to form specific associations with the content it was suggesting to them). The second goal was for our survey participants to understand the types of behaviors that we were asking them about (for example, we never used the word \enquote{strategize} in our survey). After every brainstormed set of questions we ran a \emph{pilot study} of $10$ participants. Based on the responses we got each time, we calibrated our questions (and the categories of questions more broadly) until we converged to the ones we present on \Cref{appendix:survey} (data-driven approach). 
\item Once we had converged to the set of questions, we deployed the survey on $\MTurk$ and solicited $100$ full responses. While the survey was running, we made no changes to the set of questions. We chose to deploy our survey on $\MTurk$ since the biases and the demographics of the population of workers have been well-documented in the literature~(see e.g., \cite{ipeirotis2010demographics,hitlin2016research, difallah2018demographics} for case studies).
\item In the third phase, we did qualitative data analysis using standard methods (see e.g., \cite{kuckartz2019qualitative}), which we outline next. All quantitative data (i.e., demographics) is reported with no preprocessing. For the responses that were in free-form text, we inductively created specific \emph{codes} that represented the common points in the participants' responses but were abstract enough so that they could include multiple responses. This coding step was required in order to obtain aggregate statistics from free-form text responses to our survey. The codes and their explanations for the different categories of questions can be found in \Cref{appendix:codes}.
\end{enumerate}

\subsection{Designing the Survey Questions}

One of our first steps in building the survey was deciding how the questions would be organized in blocks with shared goals. 

The first block of questions addresses the participants' usage of the platform and the time they have spent on the platform since its adoption. The purpose of this block was to assess whether our respondents spend enough time on the platform so as to have started building folk theories about how the algorithm categorizes them and decides which content to serve to them. 

The second block of questions asks the users about the types of content that they usually see on TikTok. We asked the participants both for the categories of content that TikTok puts more frequently into their feed, and the specific subcategories which they were mostly interested in. Our goal here was to make the participants to start thinking about the positive associations that the algorithm may be building with types of content that they are interested in and types of content that it puts into their feed. 

The third block of questions asks users to report the parts of their identity that were not well represented by the types of content that TikTok suggested. Then, we solicited free-form text responses regarding their best explanation for why this happened. Our goal here was to have the participants start thinking about whether \emph{they} take any actions to curate their feed that may have resulted in the algorithm presenting to them the type of content it currently does.

In our fourth block, we asked them whether a \enquote{private mode} on TikTok would make them change anything in the way that they interact with the platform. We again solicited free-form text responses. Our goal here is to understand whether participants are consciously stopping themselves from interacting with particular types of content out of fear that the algorithm will make associations that they do not want it to. 

The final block of questions explicitly asks whether the participants take any actions to \enquote{curate} the type of content they see and tests whether participants understand how the algorithm makes associations and categorizes people based on the content they consume. The goal here was to give them one more chance to think about their own curation efforts especially while being explicitly prompted to address these questions. The questions about associations was to give the participants specific examples of how the algorithm may pattern match between topics so that they could address it in the following question (i.e., the explanation of why they thought that the algorithm would associate these topics).

\subsection{Codes}\label{appendix:codes}

In the following, we explain our codes for free-form text responses from the survey participants to questions (19), (20), (25) (see Survey questions in Appendix~\ref{appendix:survey}). 

\subsubsection{Codes for Private Mode Question}
\begin{description}
\item[No change for no stated reason] Users would not change their behavior as they see no obvious reason to do so.
\item[No change for personalization reasons] Users would not change their behavior so as not to change their personalization of the algorithm.
\item[Change to engage with \enquote{avoided} content] Users would interact with content they don't want to be associated with normally (e.g., embarrassing, risky content).
\item[Change to engage with \enquote{feed-clogging} content] Users would interact with content they don't want to clog their feed.
\item[Change to increase exploration] Users would increase their exploration of new topics.
\item[Other] Users would change their behavior in other ways, for example they would switch to other platforms (because their experience would worsen as a result of less personalization), or they would engage more with content they already like.
\end{description}
\subsubsection{Codes for Curation Question}
The following are the codes we derived for the question on a private mode:
\begin{description}
    \item[Positive Association] Users either watch more, or for longer, videos that are similar to the types of content that they would like to see.
    \item[Negative Association] Users either watch less, or shorter, videos that are unlike the types of content that they would like to see.
    \item[Curation] Users describe that they engage in behaviors using both negative and positive association.
    \item[Categorization] Users specifically give evidence that they consume content that they do not like to not be categorized as not liking this type of content.
    \item[Explicit Feedback] Users describe explicit ways to give feedback to the algorithm: Likes, follows, \enquote{I am not interested} buttons, blocking.
    \item[No Curation] Users state that they do not curate content.
\end{description}
\subsubsection{Codes for the Reflection on Scenario Questions}
\begin{description}
    \item[Gender] Users state concretely that they would choose gendered content.
    \item[Stereotypes] Users state that there are stereotypes and associations. For example, a historic venue in the music scenario might also be used for physical activity, hence connecting to sports.
    \item[Similarity] Users state that they choose content that is similar, or opposite referring to closeness of content.
    \item[Gut Feeling] Users state that their choice was intuitive.
    \item[None] Users reiterate their choices while not giving reasons for them, give evidence that they are not responding under the stated hypothetical preferences, but their own, or in other way do not give a clear reason for their choice.
\end{description}

\section{Survey Participants Basic Statistics}\label{appendix:survey-participants-stats}

In this section, we report our survey participants' basic statistics; first regarding their demographics and their usage of the platform and subsequently, regarding the topics they are interested in and the algorithm puts on their feed more often.

\subsection{Demographics and Usage Statistics}

\begin{figure}[htbp]
\centering
    \begin{subfigure}[c]{.3\linewidth}
        \centering
        \includegraphics[width=\linewidth]{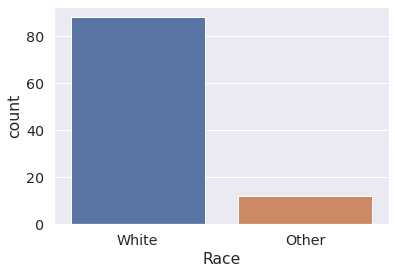}
    \end{subfigure}
    \begin{subfigure}[c]{.3\linewidth}
        \centering
        \includegraphics[width=\linewidth]{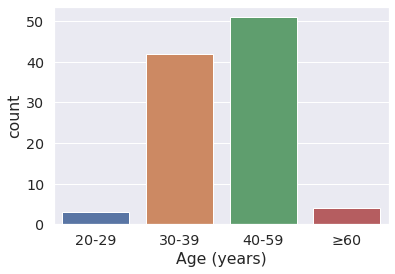}
    \end{subfigure}
    \begin{subfigure}[c]{.3\linewidth}
        \centering
        \includegraphics[width=\linewidth]{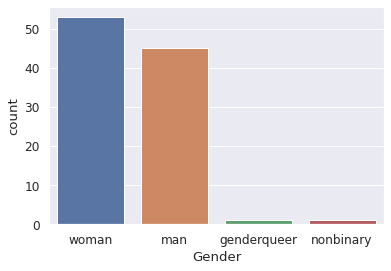}
    \end{subfigure}
   \begin{subfigure}[c]{.3\linewidth}
        \centering
        \includegraphics[width=\linewidth]{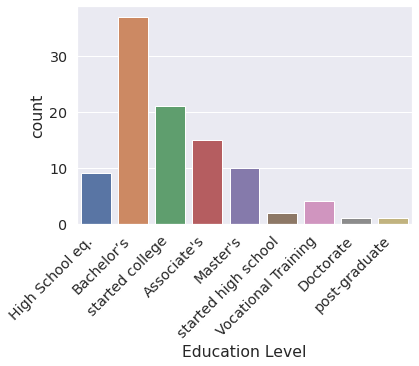}
    \end{subfigure}
    \vspace{.1\textwidth}
    \begin{subfigure}[c]{.3\linewidth}
        \centering
        \includegraphics[width=\linewidth]{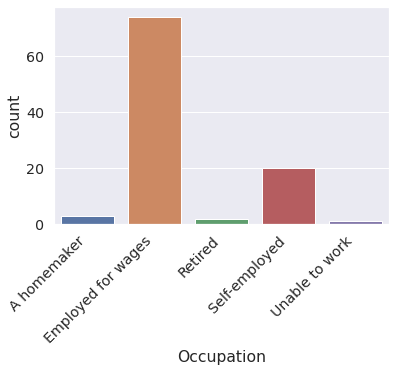}
    \end{subfigure}
    \caption{Demographics of survey participants. The $x$ axis corresponds to the categories for each plot and the $y$ axis reports the number of participants per category.}
    \label{fig:demographics}
\end{figure}

The demographics of the survey participants are shown in \Cref{fig:demographics}. Most of the participants (more than $80\%$ are white (which is in line with the population breakdown of MTurkers \cite{difallah2018demographics}). Around $50\%$ of the participants are \emph{between $40$ and $69$ years old} and more than $40\%$ are \emph{between $30$ and $39$ years old}. More than $50\%$ of the participants self-identify as \emph{women} and most of the rest self-identify as \emph{men}. We also have a small representation of folks who self-identify as \emph{genderqueer} and \emph{non-binary}. In terms of educational level and occupation, the majority of our survey participants have obtained a \emph{Bachelor's degree} and are currently \emph{employed for wages}.

\begin{figure}[htbp]
   \begin{subfigure}[c]{.3\linewidth}
        \centering
        \includegraphics[width=3.8cm]{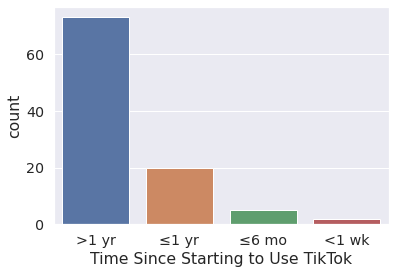}
    \end{subfigure}
    \begin{subfigure}[c]{.3\linewidth}
        \centering
        \includegraphics[width=3.8cm]{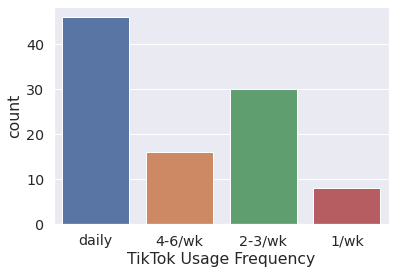}
    \end{subfigure}
    \begin{subfigure}[c]{.3\linewidth}
        \centering
        \includegraphics[width=3.8cm]{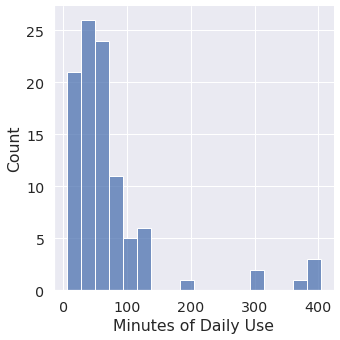}
    \end{subfigure}
    \caption{TikTok usage statistics of survey participants. The $x$ axis corresponds to the categories for each plot and the $y$ axis reports the number of participants per category.}
    \label{fig:usage-stats}
\end{figure}

The usage statistics for the survey participants are shown in \Cref{fig:usage-stats}. We see that the majority (more than $70\%$ have been using the platform for \emph{more than a year}. Around $50\%$ of the participants use the platform daily and the vast majority of all of our participants use the platform for less than $2$ hours daily. 

\subsection{Content Statistics}

In \Cref{table:content-types}, we present the different, primary content types that TikTok puts on the survey participants' feeds. We coded 34 distinct primary types of content that TikTok suggests to users, from a broad set of content types.

\begin{table}[htbp]
\centering
\begin{tabular}{lr}
\toprule
{\bf Content Types} &   {\bf Participant Count}\\
\midrule
food                    &  48 \\
funny                   &  46 \\
animals                 &  25 \\
sports                  &  20 \\
hobbies                 &  17 \\
family                  &  15 \\
dance                   &  12 \\
DIY                     &   9 \\
politics                &   9 \\
music                   &   9 \\
film \& TV               &   9 \\
fashion                 &   8 \\
home improvement        &   7 \\
money                   &   7 \\
exercise                &   6 \\
mental health           &   6 \\
science \& technology    &   6 \\
spiritual \& religion    &   5 \\
gaming                  &   5 \\
news                    &   5 \\
beauty                  &   5 \\
pop culture             &   3 \\
health                  &   3 \\
work                    &   2 \\
history                 &   2 \\
LGBTQ+                  &   2 \\
educational             &   2 \\
films \& TV              &   1 \\
travel                  &   1 \\
books                   &   1 \\
spiritualism \& religion &   1 \\
heath                   &   1 \\
conspiracy theories     &   1 \\
art                     &   1 \\
\bottomrule
\end{tabular}
\caption{Content types}\label{table:content-types}
\end{table}
\section{Selected Quotes from Survey Participants}\label{sec:quotes}

We present some interesting and informative quotes from users.

\subsection{Quotes on Representation of Identity and Curation}

\begin{displayquote}
    \emph{\enquote{I feel that Tiktok continues to put these in my feed because I almost always get sucked into watching them. That tells the algorithm that I like them, even though I am mostly just using them for background noise and have seen most of them before.}}

    \emph{\enquote{I follow a couple of people with barns, particularly horse rescues.  I often like posts not just from them but other related content that I see}}
    
    \emph{\enquote{I view this content sometimes unintentionally, but algorithmically it's recommended based on this + stuff like demographics, trends, area, etc.}}
    
    \emph{\enquote{I typically use TikTok as a way to relax and unwind, so I watch a lot of humorous content.  I think TikTok uses my watch history to fill my feed with similar videos, like the bloopers I watch regularly.}}
    
    \emph{\enquote{I was looking into finding the right form for workouts. Now tiktok probably thinks I love powerlifting.}}
    
    \emph{\enquote{TikTok puts items related to movies into my feed because they track my viewing history and also videos that I comment on and like.   There is an algorithm that runs in the background and collects this information and then sends me more of the same.}}
    
    \emph{\enquote{I think I have seen less of this on TikTok for two reasons. For the first reason, it is because there is less clean comedy on TikTok. Reason number two is that is is often hard to tell, when scrolling, which videos will have clean stand-up. So I end up watching lots of stand-up, which does a poor job of training the TikTok algorithm.}}
    
    \emph{\enquote{I believe its because I already follow or view many of these type of tiktoks so my feed is constantly showing me that type of content}}
    
    \emph{\enquote{TikTok probably puts this into my feed because it is close to the eating challenges videos.  I am interested in that, so it probably ties the two together.  Also I will watch those videos.}}
    
    \emph{\enquote{I look at this information daily in on other social media. I suspect my information is being sold. I also view on TikTok.}}
    
    \emph{\enquote{I think because I liked a video once of this type of content. I believe by me liking the video, the algorithm thought I would like to see more videos like that one.}}
    
    \emph{\enquote{I have watched/viewed this content, so it makes sense from an algorithm-based standpoint. I also believe they will randomly push videos just to have a more broad focus on varying content types for all.}}
    
    \emph{\enquote{Tiktok believes that I really enjoy jump scare videos because I may have watched some in the past, either on tiktok or on youtube. Tiktok probably sees what I have watched on all of my platforms before and curates it's suggestions based upon that. If there is some kind of pattern it sees it will seize on that and show you a bunch of videos in that genre.}}
    
    \emph{\enquote{I'm constantly looking at video game TikToks and some of those happen to be speed run related. It's easy for the algorithm to suggest that combination for my feed.}}
    
    \emph{\enquote{I interacted a lot with videos about the racial justice protests in 2020. And then I would get more academic videos about anti-carceral theory as well.}}
    
    \emph{\enquote{Because I have interacted with this content before. I have either liked this type of video or commented on this type of video.}}
    
    \emph{\enquote{I respond well and engage with funny videos. The AI learns from that.}}
    
    \emph{\enquote{I am always seeking out recipes. On most cooking videos I like and comment on the videos so I stay on that side of TikTok.}}
    
    \emph{\enquote{They put this into my feed because the algorithm picks up on when I stop and watch long form sports videos on their platform. It continues to show me this information because it knows that I like it.}}
    
    \emph{\enquote{I was probably annoyed a specific creator who I always skipped, and now it thinks I don't want any videos on the topic.}}
    
    \emph{\enquote{I have clicked not interested in this content. I also do not interact with it or watch it all the way through.}}
    
    \emph{\enquote{Perhaps I haven't trained the algorithm enough to let it know it is something I am passionate about. They probably are trying to get me to watch other things related to stuff I have watched previously.}}
    
    \emph{\enquote{i dont really view the internet very often for this specific life style choice.}}
    
    \emph{\enquote{I just do not report whether I am single or married.}}
    
    \emph{\enquote{Oh lordy, I don't know. I actually have an acute interest in religion but I prefer my religious engagement to be somewhere less frivolous than tik tok.}}

    \emph{\enquote{Again I don't look up vidoes about retail because I'm at work most of the time and don't want to see work related stuff on my free time.}}
    
    \emph{\enquote{i don't follow any tiktok hikers or anything like that. tiktok doesn't know i have an interest in it.}}
    
    \emph{\enquote{I don't watch a lot of videos or consume a bunch of content relating to exercise on social media platforms so tiktok won't have that base to figure out that that's something I like. I enjoy exercise but I do it on my own and don't post much about it.}}
    
    \emph{\enquote{I suspect because I rarely search for any travel items on TikTok, I tend to most of that on YouTube.}}
    
    \emph{\enquote{If I like a certain type of video I will make sure to like it. I will also leave a comment. Sometimes I will watch it twice so the algorithm realized I am into that kind of content.}}
    
    \emph{\enquote{I try to scroll past a video quickly if I don't like it because I don't want the algorithm to think I like this kind of content.}}
    
    \emph{\enquote{I report offensive content and try to like things I enjoy. I'm also careful which things I share.}}
    
    \emph{\enquote{I talk about topics in front of phone, google stuff I want in feed. I also like stuff just to see more of that type stuff evn though I don't like it. LIke soemtimes if my content gets to dark I try to like animal videos  and comedy more to get off the darker content for a bit.}}
    
    \emph{\enquote{Currently, I am cognizant of what category of video I think material falls under. I am careful to watch completely videos that fall under the correct category (even if I am not interested in that particular video). I am careful to skip over videos from the \enquote{wrong} categories. And I make sure to close down the app on a \enquote{wrong} category.}}
    
    \emph{\enquote{i would  try to use my own imagination  regularly as to fine tune my feed for this media site.}}
    
    \emph{\enquote{Sounds odd but sometimes I will click on something once, get out and click it again then search for it if it's a topic I like that I realize I haven't been seeing. I don't know if that actually works but it seems to.}}
    
    \emph{\enquote{I am intentional with what I search. I want stuff that serves my purpose to me to try and not clog up my fyp with stuff that I am genuinely not interested in.} (142)}
    
    \emph{\enquote{I try not to watch a video all the way through if it is something I am not interested in. Also, I try to swipe quickly on that content so as not to set off the algorithm.}}
    
    \emph{\enquote{I make sure to interact things that are specific to content types I want to see, even if I don't really love the content of that specific video.}}
\end{displayquote}

\subsection{Quotes Regarding Associations}
\begin{displayquote}
\emph{\enquote{I feel like I need to trick the system into what I do and don't want to see. I feel like if I want to see less of something, I need to change what I look at to make them realize I don't want to see certain things. Since I don't wear makeup, I shouldl be looking at more masculine things to stop from seeing makeup, for instance.}}

\emph{\enquote{There are subgroups that cluster together online according to gender association. Stereotypically feminine pursuits such as makeup do not tend to occupy the same area as sports. However, I wasn't as certain of the first, beard-cutting seems awfully sedentary for someone who wants to see sports, even if it is only of interest to males.}}

\emph{\enquote{I tried to think about how to \enquote{trick} the algorithm into going in another direction. I think choosing the options that would get me there sooner is smarter. I think a too drastic change, however, might not stick with the algorith}} [sic]
\end{displayquote}
\section{Online Survey Questions}
\label{appendix:survey}

The survey consists of six questions-blocks:

\begin{enumerate}
    \item The first block collects information regarding the respondents platform usage. Specifically, we ask them to choose from a list of pre-defined responses how often they have used TikTok in the last 30 days and when they started using the platform. For days they did use the platform, we also ask them to report how long they did.
    \item The second block consists of questions about the types of content that TikTok puts on the respondents feed \emph{most frequently}. Specifically, we first solicit answers on the top 3 types of content that the platform suggests and then we ask for each of these content types which specific subtype \emph{they} are more interested in. 
    \item The third block focuses on the accuracy of the model that TikTok has built for the respondents. We make this association of the algorithm's accuracy through the content types that are more/less frequently put on the users' feeds. Specifically, we first solicit responses regarding parts of the respondents' identities that have \emph{not} been well-represented by TikTok's algorithm. Subsequently, we ask them to give their best explanation in free-form text about why the algorithm suggests their top $3$ content types and why it is failing to represent well the parts of their identity that they declared in the beginning of the current block of questions.
    \item The fourth block of questions asks whether the respondents would change anything in their TikTok usage if TikTok offered a \enquote{private mode}. We also solicit free-form text responses with short explanations regarding their choice.
    \item The fifth block of questions focuses on the content curation from the side of the users. We first ask them to free-form text respond regarding whether they consciously take actions to curate the content they are seeing on TikTok. Subsequently, we ask them to choose from a list of predefined answers what types of content they would consume more/less of if they wanted to signal to the algorithm that they want to see more/less of a particular type of content that is related.
    \item The last block of questions elicited sociodemographic characteristics including ethnicity, education level, professional status, age, and gender. 
\end{enumerate}

For completeness, we present below the exact questions of our survey.

\subsection{Platform Usage}

\begin{enumerate}
\item How often have you used TikTok in the last 30 days?
\begin{enumerate}
\item daily
\item 4-6 times a week
\item once a week
\item less than once a week
\end{enumerate}
\item On days you use TikTok, how many minutes do you use it?
\item When did you start using TikTok?
\begin{enumerate}
\item less than a week ago
\item less than a month ago
\item less than six months ago
\item less than a year ago
\item more than a year ago
\end{enumerate}
\end{enumerate}

\subsection{Content Types}

The following questions will be about content, e.g., videos, clips, photos, memes, or text, that you watch, look at or read on TikTok.

\begin{enumerate}
\setcounter{enumi}{3}
\item Which types of content does TikTok put into your feed {\bf very frequently}? (Examples: Sports, Fashion, Food, History.) Please give three examples.
\item Which types of content does TikTok put into your feed {\bf very frequently}? (Examples: Sports, Fashion, Food, History.) Please give three examples.
\item Which types of content does TikTok put into your feed {\bf very frequently}? (Examples: Sports, Fashion, Food, History.) Please give three examples.
\item Consider the content of type  of your first example. Which more specific sub-type of content are you particularly interested in? (Examples within \enquote{sports}: workout, baseball, ballet)
\item Consider the content of type  of your second example. Which more specific sub-type of content are you particularly interested in? (Examples within \enquote{sports}: workout, baseball, ballet)
\item Consider the content of type  of your third example. Which more specific sub-type of content are you particularly interested in? (Examples within \enquote{sports}: workout, baseball, ballet)
\end{enumerate}

\subsection{Fidelity of the Recommendation System's User Model}

\begin{enumerate}
\setcounter{enumi}{9}
\item In the last 30 days, which parts of your identity were {\bf not} well-represented in what TikTok thinks you like? Give three examples.
\item In the last 30 days, which parts of your identity were {\bf not} well-represented in what TikTok thinks you like? Give three examples.
\item In the last 30 days, which parts of your identity were {\bf not} well-represented in what TikTok thinks you like? Give three examples.
\item Look at your answers for content that TikTok frequently puts into your feed. You wrote  \enquote{\texttt{their response to question (4) here}}. Give your best explanation of why TikTok puts this content into your feed. Write at least 2 sentences.
\item Look at your answers for content that TikTok frequently puts into your feed. You wrote  \enquote{\texttt{their response to question (5) here}}. Give your best explanation of why TikTok puts this content into your feed. Write at least 2 sentences.
\item Look at your answers for content that TikTok frequently puts into your feed. You wrote  \enquote{\texttt{their response to question (6) here}}. Give your best explanation of why TikTok puts this content into your feed. Write at least 2 sentences.
\item You wrote \enquote{\texttt{their response to question (10) here}} as a part of your identity that is not well-represented. Please give your best explanation for why TikTok does not show you this content.
\item You wrote \enquote{\texttt{their response question (11) here}} as a part of your identity that is not well-represented. Please give your best explanation for why TikTok does not show you this content.
\item You wrote \enquote{\texttt{their response question (12) here}} as a part of your identity that is not well-represented. Please give your best explanation for why TikTok does not show you this content.
\end{enumerate}

\subsection{Private Mode}
\begin{enumerate}
\setcounter{enumi}{18}
\item Suppose there was a \enquote{private mode} of TikTok, were you were not logged in and/or TikTok did not record which content you consume. What would you do differently on the platform, and what would you do the same way as currently? Please explain in 3-4 sentences.
\end{enumerate}

\subsection{User Content Curation}

\begin{enumerate}
\setcounter{enumi}{19}
\item Which actions do you take to curate the content you are seeing in your TikTok feed? Please describe in 1-2 sentences.
\item Assume that you want to see {\bf more sports content} in your feed in the future. Which of the following videos would you be {\bf most likely} to engage with to reach this goal?
\begin{enumerate}
\item A beard-cutting tutorial
\item A song performance in a historic venue
\item A makeup tutorial
\end{enumerate}
\item Assume that you want to see {\bf more sports content} in your feed in the future. Which of the following videos would you be {\bf most likely} to engage with to reach this goal?
\begin{enumerate}
\item A video with gossip about Kim Kardashian
\item A weight loss motivational video
\item A dog training video
\end{enumerate}
\item Assume that you want to see {\bf less makeup content} in your feed in the future. Which of the following videos would you be {\bf most likely} to engage with to reach this goal?
\begin{enumerate}
\item A dog training video
\item An NFL breakdown, play-by-play
\item A romantic comedy clip
\end{enumerate}
\item Assume that you want to see {\bf less makeup content} in your feed in the future. Which of the following videos would you be {\bf most likely} to engage with to reach this goal?
\begin{enumerate}
\item A feminist hip-hop clip
\item A weightlifting competition
\item A dance video
\end{enumerate}
\item Describe in 3 sentences why you chose to answer the previous four questions as you did.
\end{enumerate}

\subsection{Demographics}

\begin{enumerate}
\setcounter{enumi}{26}
\item What is your ethnicity?
\begin{enumerate}
\item White
\item Black of African American
\item American Indian or Alaska Native
\item Asian
\item Native Hawaiian or Pacific Islander
\item Other
\end{enumerate}
\item What is the highest education level that you have achieved?
\begin{enumerate}
\item No schooling completed
\item Nursery school to 8th grade
\item Some high school, no diploma
\item High school graduate, diploma or the equivalent (for example: GED)
\item Some college credit,  no degree
\item Trade/technical/vocational training
\item Associate degree
\item Bachelor's degree
\item Master's degree
\item Professional degree
\item Doctorate degree
\item 2-year college graduate
\item 4-year college graduate
\item graduate degree
\item post-graduate degree
\end{enumerate}
\item What is your professional status?
\begin{enumerate}
\item Employed for wages
\item Self-employed
\item Out of work and looking for work
\item Out of work but not currently looking for work
\item A homemaker
\item A student
\item Military
\item Retired
\item Unable to work
\end{enumerate}
\item What is your age?
\begin{enumerate}
    \item 20 or younger
    \item 20-29 years old
    \item 30-39 years old
    \item 40-59 years old
    \item 60 years or older
\end{enumerate}
\item Describe your gender identity
\begin{enumerate}
    \item man
    \item woman
    \item Self-describe
    \item Prefer not to say
\end{enumerate}
\end{enumerate}

\end{document}